\documentclass[11pt,a4paper]{emulateapj}
\usepackage{epsfig}
\usepackage{amsmath}
\usepackage{natbib}

\def\gs{\mathrel{\raise0.35ex\hbox{$\scriptstyle >$}\kern-0.6em\lower0.40ex\hbox{{$\scriptstyle \sim$}}}} 
\def\ls{\mathrel{\raise0.35ex\hbox{$\scriptstyle <$}\kern-0.6em\lower0.40ex\hbox{{$\scriptstyle \sim$}}}}

\def\Wm2{\,\hbox{W}\,\hbox{m}^{-2}} 
\def\gsim{\mathrel{\raise0.35ex\hbox{$\scriptstyle >$}\kern-0.6em\lower0.40ex\hbox{{$\scriptstyle \sim$}}}} 
\def\lsim{\mathrel{\raise0.35ex\hbox{$\scriptstyle <$}\kern-0.6em\lower0.40ex\hbox{{$\scriptstyle \sim$}}}} 
\def\ltsima{$\; \buildrel < \over \sim \;$} 
\def\simlt{\lower.5ex\hbox{\ltsima}} 
\def\gtsima{$\; \buildrel > \over \sim \;$} 
\def\simgt{\lower.5ex\hbox{\gtsima}}

\lefthead{Swinbank et al.}
\righthead{The high-pressure turbulent ISM in a star-forming galaxy at $z$=2.3}

\begin{document}

\title
{The ISM in distant star-forming galaxies: Turbulent pressure,
  fragmentation and cloud scaling relations in a dense gas disk at $z$\,=\,2.3}

\author{
A.\,M.\ Swinbank,\altaffilmark{1}
P.\,P.\ Papadopoulos,\altaffilmark{2}
P.\, Cox,\altaffilmark{3}
M.\ Krips,\altaffilmark{3}
R.\,J.\ Ivison,\altaffilmark{4,5}
Ian\ Smail,\altaffilmark{1}\\
A.\,P.\ Thomson,\altaffilmark{4}
R.\ Neri\altaffilmark{3}
J.\ Richard\altaffilmark{6} and
H.\ Ebeling\altaffilmark{7}
}

\setcounter{footnote}{0}

\altaffiltext{1}{Institute for Computational Cosmology, Department of Physics, Durham University, South Road, Durham DH1 3LE, UK; email: a.m.swinbank@dur.ac.uk}
\altaffiltext{2}{Max Planck Institut f\"ur Radioastronomie, Auf dem H\"ugel 69, D-53121, Germany}
\altaffiltext{3}{Institut de Radio Astronomic Millimetrique, 300 rue de la Piscine, Domaine Universitaire, 38406 Saint Martin d'Heres, France}
\altaffiltext{4}{Institute for Astronomy, University of Edinburgh, Edinburgh EH9 3HJ, UK}
\altaffiltext{5}{UK Astronomy Technology Centre, Royal Observatory, Blackford Hill, Edinburgh EH9 3HJ, UK}
\altaffiltext{6}{CRAL, Observatoire de Lyon, Universite Lyon 1, 9 Avenue Ch.\ Andre, 69561 Saint Genis Laval Cedex, France}
\altaffiltext{7}{Institute for Astronomy, 2680 Woodlawn Drive, Honolulu, HI 96822}

\begin{abstract}
  We have used the IRAM Plateau de Bure Interferometer and the Expanded
  Very Large Array to obtain a high resolution map of the CO(6--5) and
  CO(1--0) emission in the lensed, star-forming galaxy
  SMM\,J2135$-$0102 at $z=2.32$.  The kinematics of the gas are well
  described by a model of a rotationally-supported disk with an
  inclination-corrected rotation speed, $v_{\rm rot} = 320\pm
  25$\,km\,s$^{-1}$, a ratio of rotational- to dispersion- support of
  $v/\sigma = 3.5\pm 0.2$ and a dynamical mass of $(6.0\pm 0.5)\times
  10^{10}$\,M$_{\odot}$ within a radius of 2.5\,kpc.  The disk has a
  Toomre parameter, $Q = 0.50\pm 0.15$, suggesting the gas will rapidly
  fragment into massive clumps on scales of $L_{\rm J}\sim 400$\,pc.
  We identify star-forming regions on these scales and show that they
  are $\sim 10\times$ denser than those in quiescent environments in
  local galaxies, and significantly offset from the local molecular
  cloud scaling relations (Larson's relations).  The large offset
  compared to local molecular cloud linewidth-size scaling relations
  imply that supersonic turbulence should remain dominant on scales
  $\sim $100$\times$ smaller than in the kinematically quiescent ISM of
  the Milky Way, while the molecular gas in SMM\,J2135 is expected to
  be $\sim $50$\times $denser than that in the Milky Way {\it on all
    scales.}  This is most likely due to the high external hydrostatic
  pressure we measure for the interstellar medium (ISM), $P_{\rm
    tot}/k_{\rm B}\sim (2\pm 1)\times 10^{7}$\,K\,cm$^{-3}$.  In such
  highly turbulent ISM, the subsonic regions of gravitational collapse
  (and star-formation) will be characterised by much higher critical
  densities, n$_{crit}>$=10$^8$\,cm$^{-3}$, a factor
  $\gsim$1000$\times$ more than the quiescent ISM of the Milky Way.

\end{abstract}


\section{Introduction}

Observations of the molecular gas in galaxies provides important
insights into the physics of star formation.  Since all stars form out
of molecular gas -- either in self-gravitating giant molecular clouds
(GMCs) in disks such as the Milky Way or distributed in continuous
gaseous disks in ultra-luminous infrared galaxies
\citep[e.g. ULIRGs;][]{Downes98} -- studying its thermal and kinematic
state is a prerequisite for understanding star formation through-out
the Universe.  Such observations of the interstellar medium (ISM),
especially in the most extreme conditions like those in the high
star-formation density environments of local ULIRGs, can provide
powerful tests of star-formation theories and the initial conditions of
the stellar initial mass function (IMF).  In the most energetic
systems, strong star-formation feedback, high cosmic-ray energy
densities and supernovae-driven shock heating of very turbulent
molecular gas allows us to explore uncharted parameter space in current
star-formation schemes \citep{Krumholz05,PPP11}.


Given that most of the star formation in the most massive galaxies is
claimed to have occurred at $z\sim 1$--3, an era when the gas accretion
rate and molecular gas fractions of galaxies appear to be substantially
higher than today \citep{Tacconi10,Daddi10}, examining the physical and
dynamical states of the molecular gas in a star-forming system during
that era acquires special importance.  Evidence is accumulating that
the dominant mode of star formation may be very different in early
systems than that found in most local disks \citep[e.g.][]{Bournaud09,
  Jones10}.  Rather than forming stars within GMCs that condense out of
a stable disk, star formation could be triggered by fragmentation of
dynamically unstable gas-rich disks.  The resulting regions of
high-pressure ISM may be conducive to the formation of massive globular
clusters, as first postulated by \citet{Harris94}, a model which has
since been developed extensively \citep[e.g.][]{Elmegreen97,
  Schweizer96, Kratsov05}.  Such a star-formation mode likely occurs
locally, albeit on more compact scales, in the most extreme
merger-driven starbursts \citep{Downes98}.  A spatial resolution of
$\la 100$\,pc is necessary to resolve the expected size of
gravitationally unstable regions in gas-rich, star-forming disks, and
to probe the masses, sizes and hence densities of the molecular ISM on
these scales.  For example, the estimated size of the two gas-rich,
star-forming disks of Arp\,220 -- an archetypal ULIRG in the local
Universe -- is $\sim 100$\,pc, as only recently revealed by
sub-millimetre interferometry \citep{Sakamoto08}.  Achieving this
spatial resolution within starburst galaxies at $z\sim 2$ ($\sim 200
\times$ more distant than Arp\,220), with sufficient sensitivity to
measure the properties of the molecular ISM, is one of the key science
drivers for the Atacama Large Millimetre Array (ALMA).  However, prior
to full science operation of ALMA, progress can still be made by
exploiting rare examples of high-redshift starburst galaxies which have
been fortuitously strongly gravitationally lensed by foreground
clusters, boosting both the apparent size and flux of the galaxy and so
allowing imaging spectroscopy at high spatial resolution with current
millimetre arrays.

The recent discovery of a gas-rich starburst galaxy at $z=2.3$,
SMM\,J213511.60$-$010252.0 (hereafter SMM\,J2135;
\citealt{Swinbank10Nature}), provides just such an opportunity.  The
apparent brightness of SMM\,J2135 is due to a factor $\sim$35$\times$
magnification by the foreground, massive galaxy cluster,
MACS\,J2135$-$0102 \citep{Ebeling01}.  This has allowed very detailed
studies of the distribution and intensity of star formation and the
chemical make-up of the ISM in this example of the cosmologically
important high-redshift ULIRG population
(\citealt{Ivison10eyelash,Danielson11}).  In particular,
high-resolution interferometry with the Smithsonian Sub-millimeter
Array (SMA) resolved rest-frame 260-$\mu$m continuum emission in two
mirror-image sets of four clumps straddling the critical line.  The
clumps have physical (source-plane) sizes of $\sim 100$--200\,pc and
are distributed over a region $\sim 2$\,kpc within the background
galaxy.

If we can similarly resolve the ISM on 100--200\,pc scales, this would
allows -- for the first time -- a direct comparison of the scaling
relations found in local molecular clouds: the so-called Larson
relations \citep{Larson81}.  These relate the CO velocity line width of
clouds ($\sigma$) with their physical extent, $R$ ($\sigma\propto R^p$,
$p\sim 1/2$) and mean molecular gas density with size ($\langle n({\rm
  H}_2)\rangle \propto R^{-k}$, $k\sim 1$). These relations have been
found to be valid throughout the Milky Way GMCs as well as in those
extra-galactic environments where high-resolution studies of GMCs have
been possible \citep{Bolatto06}.  In particular, the universality of
the $\sigma$--$R$ relation within star-forming GMCs points towards
large-scale driving mechanisms, rather than small-scale turbulent
energy injection from within molecular clouds, as the origin of
interstellar turbulence.  The latter is a driver of the properties of
star-forming clouds \citep{Heyer04} and may define the star-formation
efficiency and the resulting stellar Initial Mass Function (IMF)
\citep{Klessen04, Larson05, Jappsen05, Klessen07}.  In this context,
Larson's relations provide a unique probe of the dynamical state of the
turbulent molecular gas in extra-galactic star-forming systems, which
are well-calibrated locally and are complementary to molecular line
ratios that constrain the density and temperature (i.e.\ the thermal
state) in these environments.

%
%
\begin{figure*}
  \centerline{
    \psfig{file=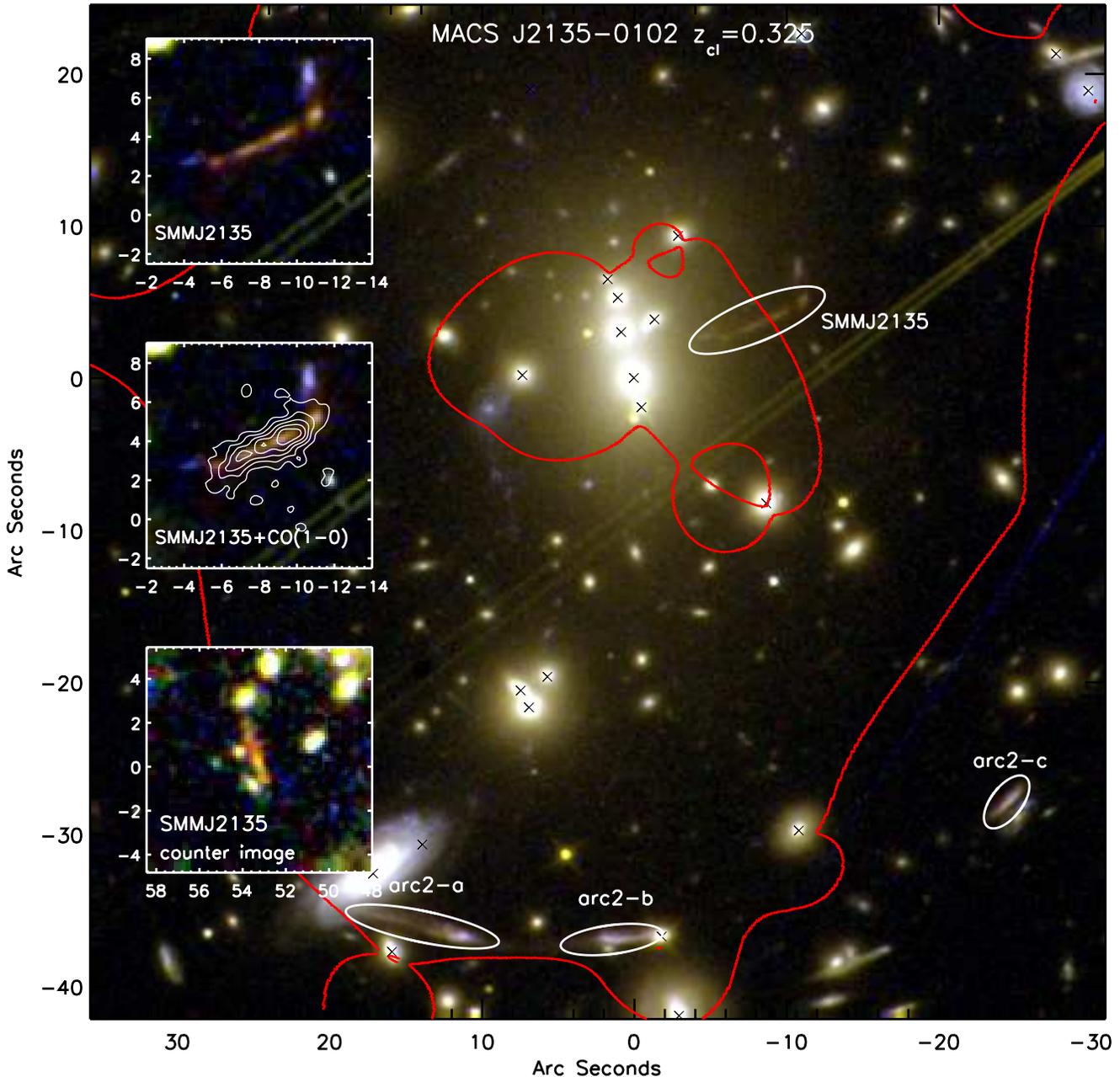,width=7in,angle=0}}
  \caption{{\it Hubble Space Telescope (HST)} ACS/WF3
    $V_{606}J_{110}H_{140}$-band image of the massive cluster
    MACS\,J2135-0102.  We overlay the $z=$2.32 critical curve from the
    best-fit lensing model (red solid curve), and also identify the
    image of SMM\,J2135-0102, as well as the triply-imaged galaxy at
    $z=$2.3 (labelled arc2-a/b/c).  The crosses show the cluster
    galaxies which are included in the lens modelling.  The insets
    show: {\it top:} a zoomed image of SMM\,2135-0102, {\it middle:}
    zoomed image of SMM\,J2135 with the CO(1--0) emission from the
    integrated cube as contours (marked at 3,6,9...$\sigma$); and {\it
      bottom:} the counter-image of SMM\,J2135-0102 located
    approximately 53$''$ due East of the Brightest Cluster Galaxy
    (BCG).  The co-ordinates for the image (and insets) are centered on
    the BCG at $\alpha:$21\,35\,12.12 $\delta:$-01\,02\,58.80 with
    North Up and East Left. }
\label{fig:hst}
\end{figure*}

%
%
\begin{figure*}
  \centerline{
    \psfig{file=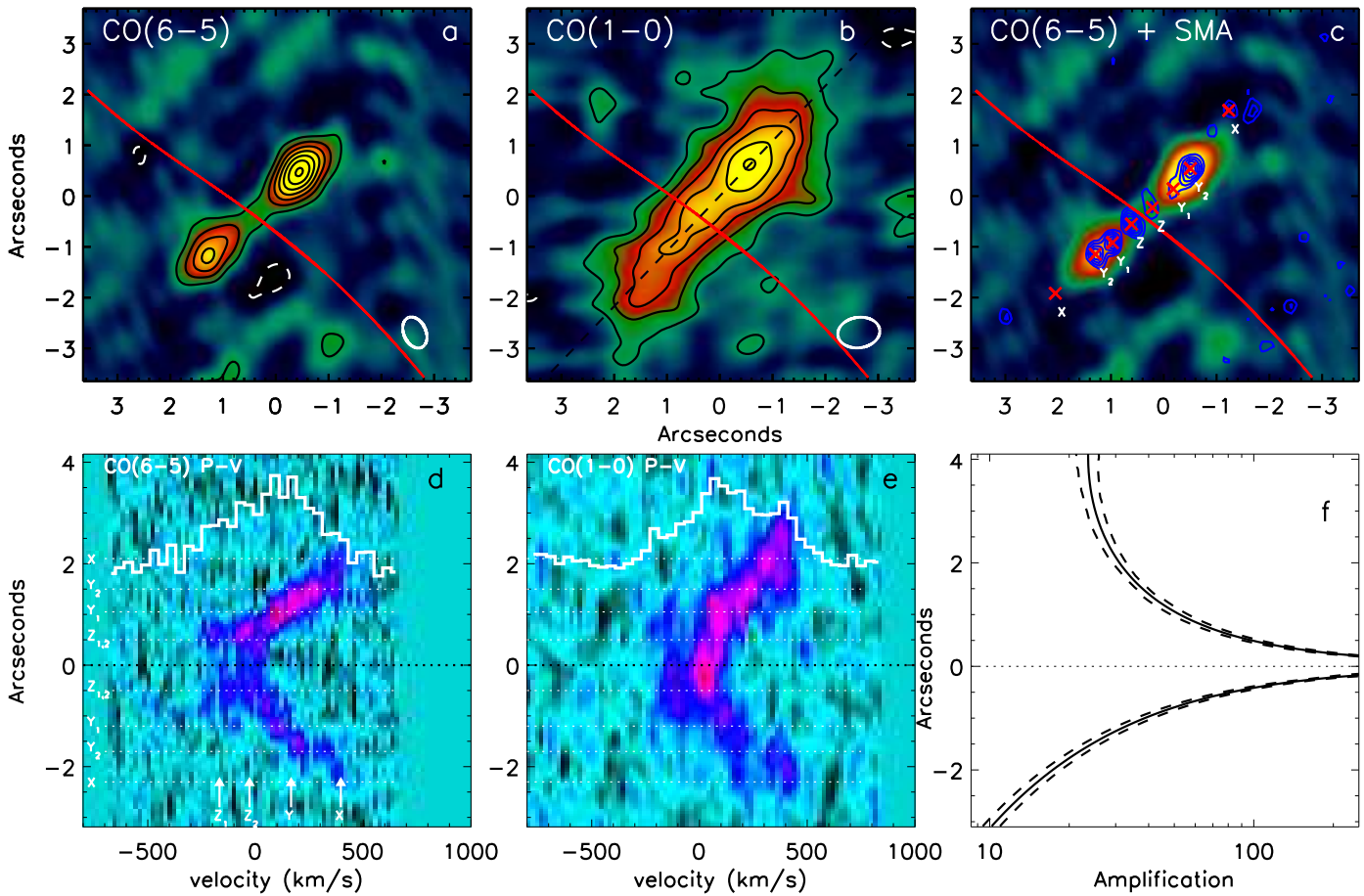,width=7.5in,angle=0}}
  \caption{Velocity-integrated image-plane maps of the {\it (a)} CO(6--5) and {\it
      (b)} CO(1--0) emission in SMM\,J2135 from our observations
    with PdBI and EVLA respectively.  In both panels, the contours
    start at 3\,$\sigma$ and are spaced by 3\,$\sigma$ thereafter, with
    negative contours at the same levels denoted by the dashed lines.
    The synthesised beam for each observation is shown, bottom right,
    in each panel, and the panels are centered at $\alpha$:
    21:35:11.638 $\delta$: -01:02:52.39 with North up and East left.
    The red solid line denotes the $z=$\,2.32 critical curve from the
    best-fit lens model.  Both CO(6--5) and CO(1--0) are extended on
    spatial scales of $\sim $\,6\,arcsec, with the emission mirrored
    across the critical curve. 
    {\it (c)}: the image-plane CO(6--5) emission-line morphology, with
    the rest-frame 260-$\mu$m continuum from our SMA observations
    overlaid as contours \citep{Swinbank10Nature}.  {\it d \& e}:
    image-plane position-velocity diagram of the CO(6--5) and CO(1--0)
    line emission extracted from the major axis at a PA 45$^{\circ}$
    east of north (as highlighted by the dashed curve in panel {(\it
      b)}.  These clearly shows velocity gradients of $\sim
    $\,500\,km\,s$^{-1}$ across $\sim $\,6\,arcsec.  The histograms
    show the collapsed, one-dimensional spectra.  T horizontal dotted
    lines denote the approximate positions of the star-forming regions
    seen in the rest-frame 260$\mu$m continuum emission, whilst the
    arrows (and labels) denote the velocities of the four kinematic
    components identified by \citet{Danielson11}.  {\it (f)} shows the
    lens amplification profile across the galaxy image for the major
    axis cross-section defined by the black dashed line shown in panel
    (b).  This plot shows that the south-eastern image of the lensed
    image is less highly amplified than the north-western image (by an
    average factor $\sim$1.4), which gives rise to the observed flux
    ratio of the northern/southern images in the sky-plane.  }
\label{fig:maps}
\end{figure*}

\begin{figure*}
\centerline{
  \psfig{figure=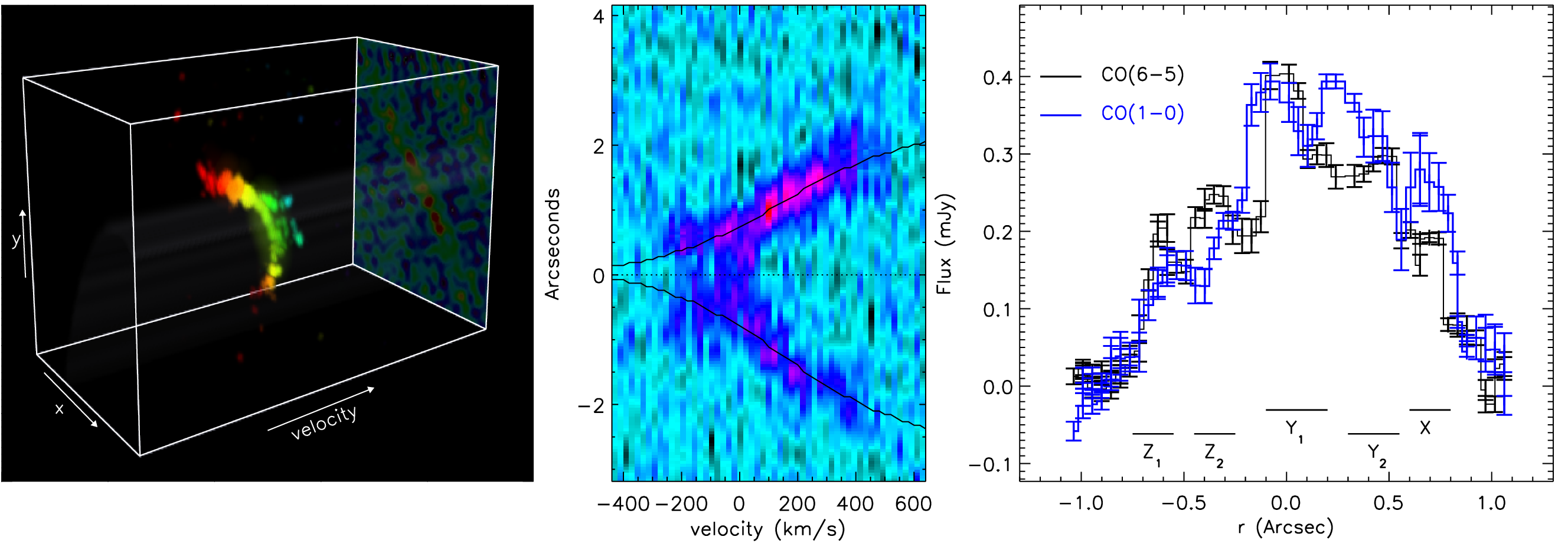,width=7in,angle=0}}
\hspace{1cm}
\caption{{\it Left:} Volume-rendered image plane cube of the CO(6-5)
  emission.  The cube is colour coded by velocity, and volume rendered
  according to the intensity in each pixel.  The 'fold' in the velocity
  field denotes the location of the critical curve where the image is
  mirrored.  The SMA 870$\mu$m (rest-frame 260$\mu$m) is projected on
  to the back face of the cube and we also include a faint grey surface
  which denotes the critical curve from the best fit lens model.
  Although the velocity integrated map in Fig.~\ref{fig:maps} appears
  relatively smooth, this rendered image clearly suggests that when
  viewed in velocity space, the gas has a ``clumpy'' morphology.  The
  full movie of SMM\,J2135-0102 can be downloaded from
  http://astro.dur.ac.uk/$\sim$ams/PdBImovie/ {\it Middle} Position
  velocity diagram for CO(6--5) extracted from the major kinematic axis
  (PA 45$^{\circ}$ east of north, as in Fig.~\ref{fig:maps}).  Here, we
  overlay the model rotation curve on both mirror images, and use this
  to extract a luminosity profile of the CO.  {\it Right:} The CO
  luminosity profile extracted from the PV diagram to emphasise the
  clumpy morphology of the CO.  So that the structure in the CO can be
  seen against the noise, the error-bars are generated by converting
  the CO flux density into number of photons detected per channel
  (accounting for the efficiency of the telescope).  We show both the
  CO(6--5) (black) and the CO(1--0) (blue) (a detailed comparison
  between the cold molecular gas seen through the CO(6--5) and
  CO(1--0), radio and sub-mm emission at high resolution will be
  presented in a forthcoming paper; R. Ivison 2011 in prep).  We also
  highlight the positions of the SMA clumps as solid bars which shows
  that the high star-formation density regions seen at rest-frame
  260$\mu$m coincides with the clumps observed in the cold molecular
  gas.}
\label{fig:PV}
\end{figure*}

In order to understand the physical processes that drive the apparently
high star-formation rates within compact star-forming regions at high
redshifts, we have obtained high-spatial-resolution observations of the
molecular gas in SMM\,J2135 traced by the CO(6--5) emission with the
IRAM Plateau de Bure Interferometer (PdBI).  We use this
spatially-resolved CO spectroscopy to probe the dynamics of the galaxy,
and the luminosities and dynamics of the star-forming gas.  We
independently confirm our results using a comparable resolution map of
the CO(1--0) data taken with the Expanded Very Large Array (EVLA).  In
\S2 we detail the observations and data analysis.  In \S3 we discuss
the properties of individual star-forming regions.  In \S4 we discuss
the wider context of our results for star-formation theories, and we
present our conclusions in \S5. Throughout the paper we use a
$\Lambda$CDM cosmology with H$_0$=72\,km\,s$^{-1}$\,Mpc$^{-1}$,
$\Omega_m$=0.27 and $\Omega_{\Lambda}$=1-$\Omega_m$ \citep{Spergel04}

\section{Observations}

We obtained {\it Hubble Space Telescope (HST)} WF3 F110W- and F160W-
band images (each of exposure time 1.2\,ks) on 2010 August 23 as part
of our snapshot program GO\# 12166; PI: H. Ebeling).  The images
comprise three 400\,s exposures dithered with a 3$''$ spacing.  The
data were reduced using {\sc multidrizzle} to provide an image with
0.12$''$ sampling and good cosmetic properties.  A visual inspection of
the image clearly identifies the red, radial image of SMM\,J2135
located precisely at the position identified by the low resolution
discovery- and high-resolution follow-up- observations at $\alpha:$
21:35:11.6 $\delta:$ -01:02:52.0 (J2000) with a spatial extent of
$\sim$7.5$''$.  A search of the WF3 image also reconfirms the
counter-image of SMM\,J2135-0102 at $\alpha:$ 21:35:15.6 $\delta:$
-01:03:13.12 (J2000).  In Fig.~\ref{fig:hst} we show the observed
$VJH$-band colour colour image around the cluster core (see
\citealt{Smail07} for a description of the {\it HST} $V$-band
observations).  We also highlight the position of a second $z=2.3$
triply-imaged system (labelled arc2-a/b/c) located $\sim$37$''$ due
South of the cluster core.  We will return to a discussion of this
imaging data, in particular in the context of the lens modelling, in
\S~\ref{sec:lensmodel}.

The redshifted CO(6--5) emission from SMM\,J2135 was observed with the
six-element IRAM Plateau de Bure Interferometer (PdBI) at 207.90\,GHz
in ``A'' configuration.  We obtained a total on-source observing time
of 6\,hr on 2010 January 6$^{th}$.  In this configuration the
synthesised beam for natural weighting is $0.67\times 0.43$\,arcsec, at
a position angle (PA) of 24.5$^{\circ}$.  The spectral correlator was
adjusted to give a frequency resolution of 2.5\,MHz across the 980-MHz
bandwidth.  The overall flux density scale was set on MWC349, with
observations of 2134+004 for phase and amplitude calibration.  Receiver
bandpass (RF) calibration was performed on 1749+096.  The data were
calibrated, mapped and analysed in the {\sc gildas} software package.
In the map, the rms is 1.5\,mJy per 25-km\,s$^{-1}$ channel.  Fitting
beams to the velocity integrated cube, we measure a velocity integrated
flux of 19$\pm$1\,Jy\,km\,s$^{-1}$, thus recovering $>$\,90\ per cent
of the single dish flux \citep{Danielson11}.

In addition, the redshifted CO(1--0) line emission was observed with
the EVLA between 2010 September and 2011 February, as part of the Open
Shared Risk Observing (OSRO) period.  We programmed the WIDAR
correlator to return two sub-bands of $64\times 2$-MHz
dual-polarisation channels.  To avoid centering the line in the noisy
channels at the edge of each sub-band, we shifted the centroid into the
40$^{\rm th}$ channel of the lower sub-band, and allowed for 10
channels of overlap between the sub-bands, yielding a total
dual-polarisation bandwidth of 236\,MHz, or 2,040\,km\,s$^{-1}$ of
velocity coverage across the line.

As a result of dynamical scheduling, data were obtained in excellent
weather conditions: yielding seven 2- or 3-hr blocks, totaling 10 and
5\,hr in the hybrid DnC and CnB configurations, respectively, although
with a similar data volume in each case.  During the blocks we switched
every few minutes between the target and the nearby, unusually bright
(4--5\,Jy), compact calibrator, 2136+004, in order to track changes in
phase and amplitude, and to determine the spectral response (bandpass).
3C\,48 was used to calibrate the flux density scale.

Typically 16 receivers were operational during a CnB block after
on-line and manual flagging.  These data were processed using standard
{$\cal AIPS\/$} procedures, as outlined by \citet{Ivison11EVLA}.
Imaging of the DnC and CnB data was accomplished using {\sc imagr},
yielding a $\sim 0.81\times 0.60$-arcsec synthesised beam (for {\sc
  robust}=5) at PA $= 84^{\circ}$.  In the map, the rms is 0.13\,mJy
per 25-km\,s$^{-1}$ channel.  In this cube, we measure a velocity
integrated CO(1--0) line flux of of 2.1\,Jy\,km\,s$^{-1}$, thus
recovering $>$\,98 per cent of the peak single-dish flux density
\citep{Swinbank10Nature}.

\section{Analysis and Discussion}

\subsection{Lens modelling}
\label{sec:lensmodel}

Since a necessary component in interpreting the CO kinematics of
SMM\,J2135-0102 is the reconstruction of the source, we first review
the lensing model constraints.  The primary constraints on the strong
lensing model for MACS\,J2135-0102 are the redshift and location of
the three images of SMM\,J2135-0102 as well as those of another
spectroscopically confirmed triply imaged galaxy at $z=2.3$ located
approximately 37$''$ due south of the brightest cluster galaxy
(labelled arc2-a/b/c in Fig.~\ref{fig:hst}).  A lensing model for the
cluster was presented in \citet{Swinbank10Nature}.  However, with the
spatially resolved spectroscopy provided by our new observations, we
can improve the strong lensing constraints using the kinematics of
SMM\,J2135-0102 to precisely locate the $z$=2.32 critical curve.

We begin by constructing the velocity-integrated emission line maps
from the CO(1--0) and CO(6-5) observations, and as Fig.~\ref{fig:maps}
shows, which are extended on scales of 4--6$''$ (see also
Fig.~\ref{fig:PV}).  We also overlay the observed 870-$\mu$m continuum
emission from \cite{Swinbank10Nature} which highlights that the
positions of the brightest regions in the sub-millimetre continuum
appear to broadly align with the molecular gas emission.  To
investigate the gas dynamics of the galaxy, we extract
position--velocity (PV) diagrams from the major axis of the image-plane
CO(6--5) and CO(1--0) cubes (at a position angle, PA = 45$^{\circ}$,
east of north as highlighted by the dashed line in Fig.~\ref{fig:maps})
and show these in Fig.~\ref{fig:PV}.  These PV diagrams show that the
galaxy has a velocity gradient of $\sim 500$\,km\,s$^{-1}$ across
$\sim$\,4--6\,arcsec which is folded (mirrored) about the critical
curve.  The sharp rise and flattening of the CO emission in the PV
diagram has the familiar kinematic signature of a rotating system, and
we discuss the dynamical properties in \S~\ref{sec:maps}.  The dynamics
of the galaxy in the image plane allow us to kinematically locate the
position of the critical curve, and by folding and cross correlating
the velocity field with itself, we kinematically locate it
0.3$\pm$0.1$''$ further south than first chosen by
\citet{Swinbank10Nature} (where it was identified from high resolution
imaging alone).  We therefore update the lensing model accordingly and
briefly discuss the lens model here.

The cluster lensing model for MACS\,J2135-0102 is derived using the
{\sc lenstool} code \citep{Kneib96,Jullo07} which employs a parametric
model of the mass distribution \citep{Richard09,Richard10}.  We use a
model with a single cluster-scale mass component (dark matter halo), as
well as individual galaxy-scale mass components centered on each
cluster member (selected from their $V-I$ colours and shown on
Fig.~\ref{fig:hst}).  We parametrise the galaxy-scale mass components
as truncated pseudo-isothermal elliptical mass distributions, each
described by the following parameters: $\{x,y,\epsilon,\theta,r_{\rm
  core},r_{\rm cut},v_{\rm disp}\}$ \citep{Kassiola93,Kneib96} which
denote the the [x/y] center, ellipticity, position angle, core radius,
cut-off radius and velocity dispersion respectively.  To reduce the
number of free parameters and match the constraints above, the
geometrical parameters ($x,y,{\epsilon},{\theta}$) describing the
cluster galaxies (including the brightest cluster galaxy) are matched
to their 2-dimensional light distributions.  We assume that the cluster
galaxies follow a scaling relation with constant mass-to-light ratio
according to an L$^{*}$ cluster galaxy, and their respective $r_{\rm
  core}$ and $r_{\rm cut}$ parameters are scales accordingly
\citep{Richard10}.  Turning to the parameters describing the
cluster-scale mass, we fix $r_{\rm cut}$ at $1\,{\rm Mpc}$ which is
motivated by its very small influence on the location of the critical
line at large values. This leaves [x,y] $\epsilon$, $\theta$,
$r_{core}$ and $v_{disp}$ as variables.

The primary constraints in defining an acceptable strong lens model for
cluster lensed are that it reproduces the locations and redshift of the
spectroscopically confirmed triple-image lens systems
\citep[e.g.][]{Smith09}, and in the case of MACS\,J2135-0102 we also
demand that it precisely predicts the location of the critical curve.
The best-fit model (Table~1) obtains a fit with an r.m.s. of 0.15$''$
between the predicted and observed positions of the multiple images.
Since the {\sc lenstool} code incorporates a Markov Chain Monte Carlo
(MCMC) sampler, we also generate 1000 error models that lie within the
$68\%$ confidence intervals around the best fit parameters (after
marginalising over the other parameters in each case), and construct
the amplification maps in each case.  In the best fit model, the
cluster-scale component is centered South East of of the BCG (Table~1),
and the enclosed mass within an aperture of 250\,kpc is
M=3.3$\pm$0.3$\times$10$^{14}$\,M$_{\odot}$ with an Einstein radius of
$\theta_E$=34.5$\pm$2.0$''$ for $z$=2.32 (Richard et al. 2011 in prep).
In Fig.~\ref{fig:hst}\&~\ref{fig:maps} we show the critical curve on
the {\it HST} and CO imaging, as well as amplification as a function of
position across the long axis of the arc (and the associated 1$\sigma$
error from the family of models which adequately describe the lensing
mass).  The updated model suggests a luminosity weighted amplification
of the system of $\mu$=37.5$\pm$4.5 which higher, but within 1$\sigma$
of the previous estimate of 32.5$\pm$4.5 from \citet{Swinbank10Nature}.

\begin{table*}
{\scriptsize
\begin{center}
{\centerline{\sc Gravitational Lens Model Parameters}}
\smallskip
\begin{tabular}{lccccccc}
\hline
\hline
\noalign{\smallskip}
                 & ${\Delta}{\rm RA}$ & ${\Delta}{\rm Dec}$ & $\epsilon$         & $\theta$           & $r_{\rm core}$    & $r_{\rm cut}$  & $v_{\rm disp}$  \\
                 & ($''$)             & ($''$)            &                      &  (deg)             &   (kpc)          &  (kpc)       & (km/s)         \\
\hline
DM halo          & -0.7$\pm$0.8      &  -2.5$\pm$0.5     &  0.25$\pm$0.10       & -20$\pm$2         &  83$\pm$8        &   [1000]     &  1190$\pm$30   \\
BCG              & [$0.0$]            & [0.0]             &  [0.15]              & [148]              &  [0.2]           &   158$\pm$8  &  262$\pm$7     \\
L$^{*}$ galaxies & -                  & -                 &  -                   & -                   & [0.15]           &  [45]       &  179$\pm$18     \\
\hline\hline
\end{tabular}
\caption{{\footnotesize Notes: Numbers in square brackets are fixed in
    the fit.  Position angles defined such that positive is East of
    North.  For a complete description on the choice of these
    parameters, see \citet{Richard09}.}
}
\end{center}
}
\end{table*}

\subsection{Gas distribution and dynamical properties}
\label{sec:maps}

\begin{figure}
\centerline{
  \psfig{figure=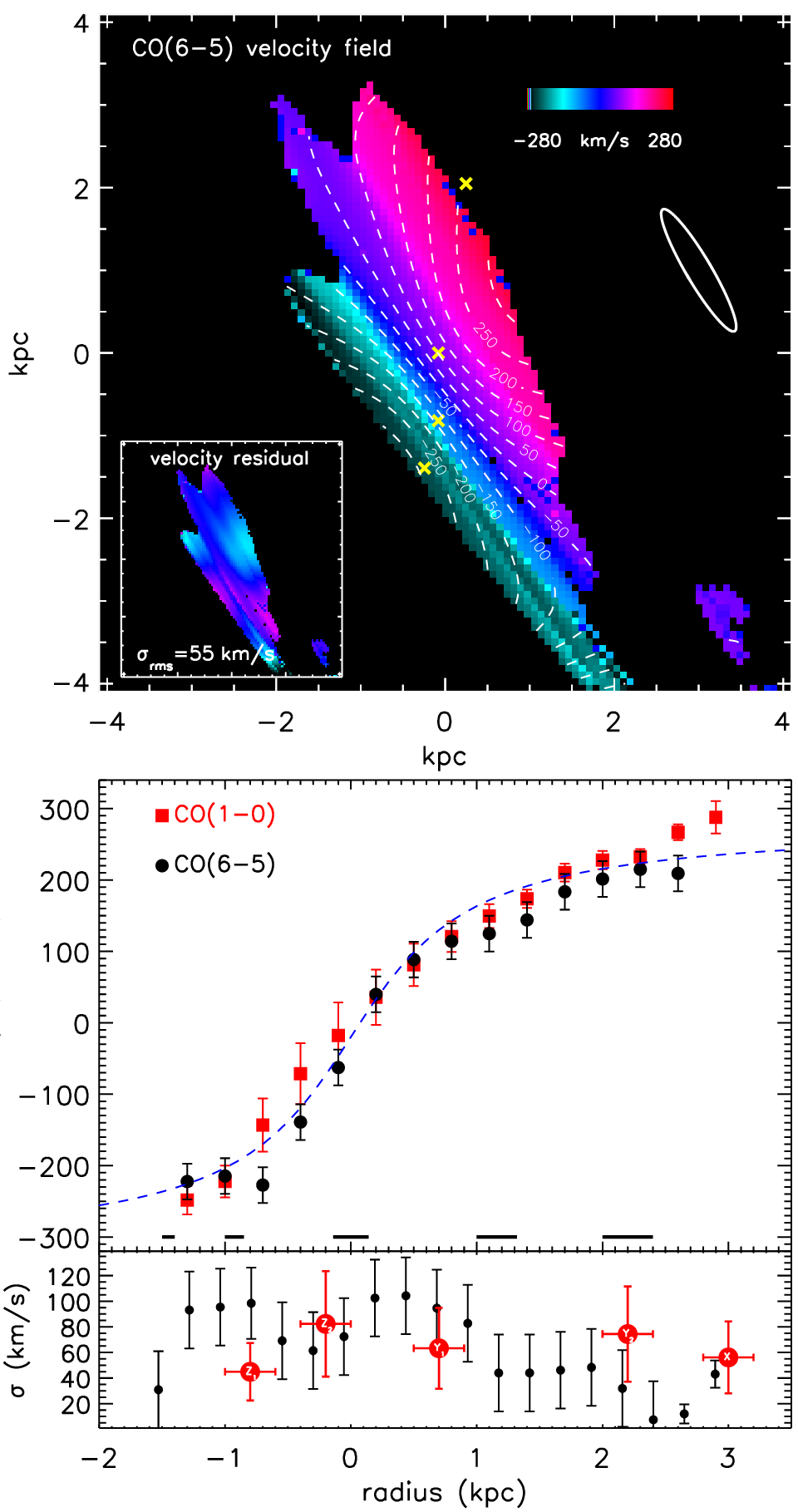,width=2.7in,angle=0}}
\hspace{1cm}
\caption{{\it Top:} Two-dimensional kinematics of the molecular gas in
  SMM\,J2135 from the CO(6--5) emission line after correcting for
  lensing amplification.  The velocity field shows a velocity gradient
  of $\sim$\, 500\,km\,s$^{-1}$ across $\sim $\,3\,kpc in projection,
  indicating a dynamical mass of
  (6.0\,$\pm$\,0.5)\,$\times$\,10$^{10}$\,M$_{\odot}$ (corrected for
  inclination, $i = $\,60\,$\pm$\,8$^{\circ}$).  The contours denote
  the best-fit disk model and the crosses show the locations of the
  star-forming regions seen in the rest-frame 260-$\mu$m continuum
  imaging from \citet{Swinbank10Nature}.  The source-plane beam is
  shown in the top right-hand corner and is approximately
  150\,$\times$\,850\,pc on average across the galaxy image, but
  reaches $\sim$\,90\,pc in the most highly amplified region.  {\it
    Inset:} Velocity residual after subtracting the best-fit disk
  model.  {\it Bottom:} Source-plane, one-dimensional velocity- and
  velocity-dispersion- profile from the CO(6--5) emission line
  extracted from the major kinematic axis using a synthetic slit of
  width 1\,kpc.  The dashed curve denotes the rotation curve of the
  best-fit two-dimensional velocity field.  The solid horizontal bars
  denote the resolution of the CO(6--5) observations as a function of
  position within the galaxy.  The lower panel shows the
  one-dimensional velocity dispersion across the same major kinematic
  axis of the galaxy (small points) whilst the the large solid points
  show the velocity dispersion of the individual star-forming regions
  as identified from the SMA morphology.
}
\label{fig:COvel}
\end{figure}

To construct a two-dimensional map of the velocity field from the CO
emission in the source plane we use the best-fit lensing model to
calculate the mapping between image- and source- plane, and ray-trace
each pixel to re-construct the source-plane datacube.  We then fit the
CO(6--5) emission at each pixel with a Gaussian profile using a
$\chi^2$ minimisation procedure, accepting a fit only if the emission
is detected at $>5\sigma$.  Where a fit is made, we measure the central
velocity, line width ($\sigma$) and line flux.  The resulting
source-plane velocity field of the gas is shown in Fig.~\ref{fig:COvel}
and resembles a rotating system with a peak-to-peak velocity of $240\pm
25$\,km\,s$^{-1}$ within a radius of $\sim3$\,kpc.  In
Fig.~\ref{fig:COvel}, we also show the extracted one-dimensional
rotation curve of the CO(6--5) emission as well as the line-of-sight
velocity dispersion.

In order to estimate the disk inclination and true rotational velocity,
we model the velocity field with a rotating disk.  We use an arctan
function to describe the shape of the rotation curve such that $v(r) =
v_{\rm c}\,{\mbox arctan}(r/r_{\rm t})$, where $v_{\rm c}$ is the
asymptotic rotational velocity and $r_{\rm t}$ is the effective radius
at which the rotation curve turns over \citep{Courteau97}.  We
construct the two-dimensional kinematic model for the galaxy using six
free parameters: $v_{\rm c}$, $r_{\rm t}$, $x$, $y$ (the central
position), PA and disk inclination, $i$, and use a genetic algorithm
with $10^5$ random initial values to search for a best fit.  We demand
$>30$ generations are performed before testing for convergence to a
solution.  The best-fit kinematic model is overlaid on
Fig.~\ref{fig:COvel} and has $r_{\rm t} = 0.4\pm 0.1$\,kpc, $i = 60\pm
8^{\circ}$ and $v_{\rm t} = 275\pm 20$\,km\,s$^{-1}$, providing an
estimate of the dynamical mass of $M_{\rm dyn} = (6.0\pm 0.5)\times
10^{10}$\,M$_{\odot}$ for an inclination of $60\pm 8^{\circ}$.  We note
that the total gas mass from a large velocity gradient (LVG) analysis,
$M_{\rm gas} = 4\times 10^{10}$\,M$_{\odot}$ \citep{Danielson11}
together with the total stellar mass, $M_{\star} = (3\pm
1)\times$10$^{10}$\,M$_{\odot}$ \citep{Swinbank10Nature}, indicates
baryon-dominated dynamics within the extent of the gas disk, (M$_{\rm
  gas}$+M$_*$)/M$_{\rm dyn}$=1.1$\pm$0.2, and a gas fraction of $f_{\rm
  gas}\sim M_{\rm gas}/M_{\rm dyn}\sim M_{\rm gas}/(M_{\star} + M_{\rm
  gas})=$0.6$\pm$0.1.  This suggests a total baryonic mass surface
density (stars and gas) within the disk of $(3.0\pm 0.5)\times
10^{9}$\,M$_{\odot}$\,kpc$^{-2}$.

The ratio of rotation/dispersion within a disk provides an important
constraint on the amount of turbulence within the ISM.  Nebular
emission line studies have shown that high velocity dispersions are a
common feature of high redshift galaxies ($\sim$4--10$\times$ higher
than comparably luminous local galaxies), possibly related to the high
gas fractions and star-formation rates
\citep[e.g.][]{Lehnert09,ForsterSchreiber09,Genzel11}.  However,
measuring the typical local velocity dispersion is not trivial,
especially with low ($\gsim$\,kpc) spatial resolution, if the
velocity gradient within the beam is much larger than the local
velocity dispersion.  In this case, beam smearing can produce a large
bias in the recovered value (Davies et al.\ 2011).

Since the spatial resolution obtained by our observations is much
higher than typically acheived in high-redshift galaxy studies and the
orientation of the glaaxy is such that the maximum amplification is
aligned roughly parallel to the velocity gradient, beam smearing has
less of an effect on our intrinsic, although a small correction is
still needed.  The maximum velocity gradient in SMM\,J2135-0102 is
$dV/dr=0.15$\,km\,s$^{-1}$\,pc$^{-1}$, thus over a beam size of
$\sim$200\,pc there is a maximum contribution of 25\,km\,s$^{-1}$ to
the velocity dispersion.  To account for this, we therefore correct the
local velocity dispersion in each pixel and determine an average line
of sight velocity dispersion of 91$\pm$12\,km\,s$^{-1}$ across the
galaxy image (we note that if instead we simply measure the uncorrected
average line-of-sight velocity dispersion we derive $\sigma = 97\pm
10$\,km\,s$^{-1}$).  Thus, we derive an average inclination-corrected
$v/\sigma = 3.5\pm 0.2$.  This is larger than that found in many of the
star-forming galaxies studied through their nebular emission
\citep[e.g.][]{ForsterSchreiber09, Law09}, and similar to that found in
some local ULIRGs \citep{Downes98}.

As an independent check of the kinematics of the gas reservoir in
SMM\,J2135 we use our CO(1--0) map from the EVLA.  The resolution of
this map is slightly poorer, but still comparable to that of the PdBI
CO(6--5) observations, but importantly the CO(1--0) line emission
encompasses much cooler and less dense gas ($\rm n^{(10)} _{crit}$$\sim
$400\,cm$^{-3}$ $\rm E_{1}/k_B$=5.5\,K) associated with
non-star-forming ISM {\it in addition} to the warmer and denser gas,
typically found in star-forming regions which is traced by the CO(6--5) emission
($\rm n^{(65)} _{crit}$$\sim
$$6\times 10^4$\,cm$^{-3}$, $\rm E_{6}/k_B$$\sim $116\,K).  The
velocity field and rotation curves from CO(1--0) in Fig.~\ref{fig:maps}
show good agreement with those from CO(6--5) although it is clear that
the CO(1--0) emission extends at least 1$''$ ($\sim 1$\,kpc) further
from the critical curve in both images of the lensed galaxy (even
accounting for the slightly larger beam).  This bias: much broader and
more spatially extended low-$J$ line emission compared to high-$J$
lines has recently been reported for a number of high-redshift
sub-millimetre galaxies \citep{Ivison11EVLA,Riechers11a,Riechers11b} in
addition to low redshift starbursts
\citep[e.g.][]{PPP98,Mao00,Weiss01,Weiss05c,Walter02}.  A detailed
comparison between low- and high-$J$ CO emission and the radio
continuum emission in SMM\,J2135 on sub 100\,pc scales from
A-configuration EVLA observations will be discussed in an upcoming
paper (R. Ivison et al., in prep).

\subsection{The high-pressure molecular ISM in SMM\,J2135}

The pressure in the ISM -- overwhelmingly due to the non-thermal,
macroscopic motions in the highly supersonic, turbulent molecular gas
-- is an important parameter for understanding the formation of stars.
It effects the normalisation of the $M({\rm H}_2)$--L$_{\rm CO(1-0)}$
relation \citep{Bryant96}, Larson's relations \citep{Chieze87,
  Elmegreen89b} and the star-formation efficiency $\Delta ({\rm
  SFR})/\Delta m_{\rm gas}$ \citep{Schaye08}.  It also plays a
fundamental role in the H\,{\sc i} $\rightarrow$ H$_2$ phase transition
and thus in the formation of molecular clouds in spiral disks
\citep[e.g.][]{Elmegreen93, Honma95}, and their fragmentation into
dense cores \citep[e.g.][]{Chieze87b}.  Finally, high pressure in a
molecular inter-cloud medium may be responsible for the shock
compression of individual GMCs as they fall into the central regions of
ULIRGs and their subsequent burst of star formation \citep{Solomon97}.

Before computing the average gas pressure in the ISM in SMM\,J2135 it
is important to note that the existence of random macroscopic gas
motions does not automatically imply that the corresponding pressure
term $1/3\rho_{\rm gas} \langle \sigma^2 \rangle$ is physical and thus
dynamically important.  This requires a mechanism for momentum transfer
between the components to transmit the pressure, and one capable of doing so
 over a short ``collision'' mean-free path length with respect to the size of the
gas distribution.  Direct collisions between kinematically distinct
components may not easily satisfy these two criteria in highly
supersonic molecular clouds.  On the other hand, magnetic field lines
``threaded'' through the various gas clumps may provide a suitable
coupling between molecular clouds \citep{Elmegreen89}.  However, it
must be noted that the molecular gas in compact starbursts
(e.g.\ ULIRGs -- perhaps the closest analogues to SMM\,J2135) may be
very different from that found in the disk of the Milky Way, with a
much larger -- perhaps dominant -- gas-mass fraction at densities of
$n({\rm H}_2)\ga 10^5$\,cm$^{-3}$ \citep[e.g.][]{Gao04}.  At such high
densities and in the high-extinction environment of ULIRGs and SMGs,
the average ionisation fraction of the gas may be very low, allowing
ambipolar diffusion to very effectively remove magnetic field lines
from the gas, thereby reducing the coupling between components.
Nevertheless, for the purposes of this study we assume that the
computed gas pressure is physical and dynamically important.

The dynamics and surface density of gas and stars can be used to
estimate the ISM pressure, which can then be compared to that seen in
gas-rich galaxies locally.  For a rotating disk of gas and stars the
external hydrostatic pressure at mid-plane is given by:

\begin{equation}
  \rm P_{\rm tot} \approx \frac{\pi}{2}G \Sigma_{\rm gas}\left[\Sigma_{\rm gas}+\left(\frac{\sigma_{\rm gas}}{\sigma_{*}}\right)\Sigma_{*}\right]
  \label{eqn:Ptot}
\end{equation}

\noindent where $\Sigma_{\rm gas}$ and $\Sigma_{\star}$ are the surface
density of the gas and stars, and $\sigma_{\rm gas}$ and $\sigma_{*}$
are the vertical velocity dispersion of the gas and the stars,
respectively (with no assumption that the gas scale length is smaller
than the stars; c.f.\ \citealt{Blitz06}).  For the Milky Way this
yields $P_{\rm tot}/k_{\rm B}\sim 1.4\times 10^4$\,cm$^{-3}$\,K
\citep{Elmegreen89}.  To estimate the stellar surface density in
SMM\,J2135, we use the stellar mass estimate from
\citet{Swinbank10Nature}, $M_{\star} = (3\pm 1)\times
10^{10}$\,M$_{\odot}$ to derive $\Sigma_{\star} = (2.4\pm 0.5)\times
10^{3}$\,M$_{\odot}$\,pc$^{-2}$ (and assume that the stars and gas have
the same spatial distribution, which is reasonable given the {\it HST}
and {\it Spitzer} IRAC morphologies; Fig.~\ref{fig:hst} and
\citealt{Swinbank10Nature}), whilst the gas surface density we derived
above as $\Sigma_{\rm gas} = 3.2\times 10^{3}$\,M$_{\odot}$\,pc$^{-2}$.

The average velocity dispersion of the gas in the disk is $\sigma_{\rm
  gas} = 91\pm10$\,km\,s$^{-1}$, and we have to assume that the
velocity dispersion of the stars is comparable, (although we allow this
to vary between 50--150\,km\,s$^{-1}$ in the following results).  The
resulting mid-plane hydrostatic pressure is $P_{\rm tot}/k_{\rm B}\sim
(2\pm 1)\times 10^{7}$\,K\,cm$^{-3}$.  This pressure is $\sim 1000
\times$ higher than the typical ISM pressure in the Milky Way
($10^{4}$\,K\,cm$^{-3}$), and approximately 2--$3\times$ higher than
seen in other local, gas-rich environments, such as the Galactic Center
or the Antennae \citep{Rosolowski05,Keto86,Wilson03}.  Only the compact
disks in local ULIRGs have comparable mid-plane pressures due to their
high gas-mass surface densities ($\ga 5\times
10^3$\,M$_{\odot}$\,pc$^{-2}$, \citealt{Downes98}).

We can compare our measured ISM pressures to estimates for five
high-redshift disks galaxies from \citet{Genzel08} with
well-constrained kinematics.  Although gas masses do not exist for
these galaxies directly, they have been inferred indirectly using the
H$\alpha$ luminosity density and the Schmidt-Kennicutt relation.  This
provides an estimate of the likely gas surface density to within a
factor $\sim 3\times$ depending on the choice of parameters
(e.g.\ \citealt{Genzel10} but see also \citealt{PPP10}).

The (inferred) median gas mass for the \citet{Genzel08} sample is
$M_{\rm gas} = 0.2\times$10$^{11}$\,M$_{\odot}$, with stellar masses of
$M_{\star}\sim (0.3$--$6)\times 10^{11}$\,M$_{\odot}$ in disks of scale
length 4--7\,kpc and vertical scale heights of $h_{\star} =
0.8$--1.6\,kpc \citep{Shapiro10}.  Together, these properties suggest
ISM pressures in the range $P_{\rm tot}/k_{\rm B}\sim (0.3$--$10)\times
10^7$\,K\,cm$^{-3}$, suggesting that high gas pressures may be a
ubiquitous feature of high-redshift, gas-rich systems.

%
%
\begin{figure*}
\centerline{
  \psfig{figure=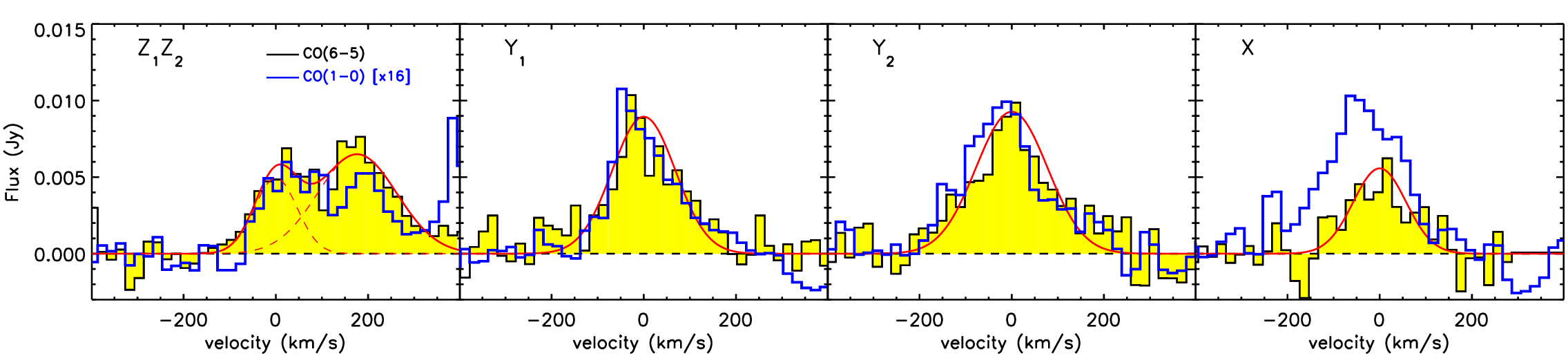,width=7.0in,angle=0}} 
\hspace{0.5cm}
\caption{The CO(6--5) and CO(1--0) spectra extracted at the positions
  of the star-forming regions identified by the peaks in the rest-frame
  260-$\mu$m emission.  The datacube has first been corrected for the
  large-scale velocity field in order to remove the broadening of the
  lines due to large scale motions.  Components Z$_1$Z$_2$, Y$_1$,
  Y$_2$ and ``X'' correspond to kinematic components ``Z'' and ``Y''
  and ``X'' in \citet{Danielson11} (and SMA clumps ``Z, Y \& X'' in
  Fig.~\ref{fig:maps}).  The clumps have velocity dispersions of
  45--85\,km\,s$^{-1}$ which is a factor $\sim 5\times$ higher than
  expected for local, quiescent GMCs given their size and luminosity.
  We also overplot the equivalent CO(1--0) line widths, which broadly
  trace the same kinematics as the CO(6--5) emission.}
\label{fig:spec_clumps}
\end{figure*}

\subsection{Disk stability and fragmentation scale}

Having establish that the ISM pressure is high, we next examine
the susceptibility of the gas disk to collapse.  In a rotating disk,
perturbations smaller than a critical wavelength, $L_{\rm max}$, are
stabilised against gravity by the velocity dispersion; those larger
than some critical wavelength, $L_{\rm min}$, are stabilised by
centrifugal force. If the dispersion and rotation velocity are too low,
$L_{\rm min}>L_{\rm max}$ and perturbations of intermediate wavelength
grow exponentially.  The Toomre parameter, $Q = L_{\rm max}/L_{\rm
  min}$, characterises the stability against local axisymmetric
perturbations of a disk supported by differential rotation and random
motion \citep{Toomre64}.  Galaxies with $Q<1$ are unstable on scales
between $L_{\rm max}$ and $L_{\rm min}$ and will fragment into giant,
dense clumps.  This could trigger star formation in clouds of much
higher mass and size than GMCs seen in local spiral galaxies with
$Q>1$.  Dynamical friction, viscosity and tidal interactions may
subsequently cause these clumps to migrate toward the center of the
galaxy potential, forming a bulge which stabilises the system against
further fragmentation \citep[e.g.][]{Genzel08}.

The Toomre parameter is calculated from
\begin{equation}
  Q = \frac{\sigma_{\rm r} \kappa}{\pi G \Sigma}
  \label{eqn:Q}
\end{equation} 
where $\sigma_{\rm r}$ is the one-dimensional random velocity
dispersion, $\kappa$ is the epicyclic frequency and $\Sigma$ is the
surface mass density \citep{Toomre64}.  The value of $\kappa$ is
uncertain as it depends on the unknown mass distribution; our
observations are consistent with a range $\sqrt{2} \frac{V_{\rm
    c}}{R}$--$2 \frac{V_{\rm c}}{R}$ corresponding to constant $V_{\rm
  c}$ and $V_{\rm c} \propto R$, respectively.  Disk thickness and gas
abundance also affect the value of $Q$, and we have assumed that the
measured velocity dispersion is equal to $\sigma_r$.  Adopting $\kappa
= \sqrt{3}\frac{V_{\rm max}}{R}$ appropriate for a uniform disk, and
using the dynamical mass to estimate the surface mass density,
$\Sigma$, we find an inclination-corrected $Q = 0.50\pm 0.15$ (the
error is contributed equally from the uncertainty in the dynamical mass
and the assumed $\kappa$).  The gas disk in SMM\,J2135 thus appears to
be dynamically unstable ($Q<1$), perhaps more gravitationally unstable
than the compact, gas-rich disks residing in ULIRGs where $Q$ is closer
to unity \citep{Downes98}.

Empirically, this instability will cause large, dense gas condensations
to form in the molecular gas distribution.  This fragmentation should
occur on the scale of the Jeans length for dispersion support.  In a
uniform disk, the largest scale for which velocity dispersion
stabilises against gravitational collapse is:
\begin{equation}
  L_{\rm J} = \frac{\pi \sigma^2}{8 G \Sigma}
  \label{eqn:Lj}
\end{equation}
which can be estimated from the dispersion and dynamical mass
density. As with $Q$, the unknown mass distribution, disk height, gas
content and directional dependence of $\sigma$ result in an uncertainty
of around 30 per cent. The resulting instability scale is $L_{\rm J} =
400\pm 150$\,pc and the corresponding unstable mass scale is
$\sigma^4/G^2\Sigma\sim 10^9$\,M$_{\odot}$, equivalent to the entire
molecular gas reservoir of a spiral disk such as the Milky Way.

Our previous high-resolution, rest-frame far-infrared imaging has shown
that the galaxy contains several bright star-forming regions, with
star-formation rates of 30--100\,M$_{\odot}$\,yr$^{-1}$ each
(equivalently, luminosities of $2$--$5\times 10^{11}$\,L$_{\odot}$),
spread across a radius of $\sim 2$\,kpc \citep{Swinbank10Nature}.
Although the velocity integrated CO(6--5) emission line maps appear
smooth, Fig.~\ref{fig:PV} shows that when the cube is viewed in the
three dimensions the CO(6--5) and CO(1--0) have complex morphologies.
Hence, to examine the structures in the gas distribution and so examine
how these might relate to the rest-frame 260$\mu$m continuum
morphology, we extract position-velocity plots and the one-dimensional
intensity profiles of the CO(6--5) and CO(1--0) emission from these PV
diagrams and also show these in Fig.~\ref{fig:PV}.  These show that the
gas morphology is indeed structured, apparently comprising (at least)
five clumps across the $\sim$5\,kpc disk.

\citet{Danielson11} demonstrated that to model the integrated
multi-transition CO kinematics of SMM\,J2135 required multiple velocity
components which they denoted X, Y, Z$_1$ and Z$_2$.  Our
high-resolution CO dynamics now allow us to spatially locate these
structures and link them to the star-forming regions seen in the
rest-frame 260-$\mu$m emission.  Fig.~\ref{fig:maps}\&~\ref{fig:PV}
show that the lowest velocity material in \citet{Danielson11},``Z$_1$''
and ``Z$_2$'' (at $\Delta v$=28$\pm$9 and -167$\pm$9\,km\,s$^{-1}$ with
respect to the systemic respectively) appear to relate to two clumps
separated by $<$\,0.2$''$ in projection on the sky, are associated with
the rest-frame 260-$\mu$m emission closest to the critical curve.
Kinematic component ``Y'' in \citet{Danielson11} ($\Delta
v$=165$\pm$13\,km\,s$^{-1}$) appears to comprise two or three gas rich
clumps between $v\sim 110$--280\,km\,s$^{-1}$.  This gas-rich component
is centered within $\sim 0.2''$ of the brightest emission at rest-frame
260\,$\mu$m.  Finally, the highest velocity material,``X'' in
\citet{Danielson11} ($\Delta v$=396$\pm$9\,km\,s$^{-1}$) corresponds to
the highest velocity, faintest emission in the rest-frame 260-$\mu$m
and CO(1--0) and CO(6--5) maps.

Given the correspondence between peaks in the molecular gas and the
rest-frame 260-$\mu$m emission, we will use the rest-frame 260-$\mu$m
map as a guide to isolate the star-forming regions, extracting the
properties of the molecular gas at these locations.  Clearly this
biases our results towards the properties of the brightest star-forming
regions, potentially missing some of the gas emission lies outside
these actively star-forming regions, however, it has the advantage that
we do not need to identify peaks directly from the CO emission -- which
itself is difficult to accomplish, without introducing biases due to
surface brightness effects and signal-to-noise.  We therefore restrict
the following analysis to the locations of the brightest regions in the
rest-frame 260-$\mu$m emission.

Before extracting the size, luminosity and velocity dispersion of the
molecular gas we must first remove the large-scale velocity structure
of the galaxy (and hence any rotational contribution to the line widths
of the clumps), as well as estimating the contribution of the velocity
gradient within the beam.  We therefore subtract the best-fit kinematic
model for the galaxy velocity field from the cubes and then extract
CO(6--5) and CO(1--0) spectra and measure the line luminosity, velocity
dispersion and central velocity (Fig.~\ref{fig:spec_clumps}).  As in
\S~\ref{sec:maps} we use the model velocity field to estimate the local
velocity gradient ($dV/dR$) within the beam and remove this in
quadrature from the velocity dispersion.  The largest corrections are
in the central regions of the galaxy where the local velocity gradient
is large (25\,km\,s$^{-1}$ over 200\,pc).  We note that in all four
regions from where we extract spectra, the CO has intrinsic velocity
dispersions of 45--82\,km\,s$^{-1}$ from CO(6--5), and that the
CO(1--0) line widths are a factor 1.2\,$\pm$\,0.1 times higher (even
after beam-matching the cubes).  However, this is as expected if the
CO(6--5) traces denser gas regions, more deeply embedded inside
molecular clouds, and thus subjected to higher average pressures.  To
estimate the approximate sizes for the molecular gas associated with
the star-forming regions, we collapse the CO(6--5) source-plane
datacube over the $\pm$2$\times$FWHM of the CO line at its systemic
velocity and fit an elliptical Gaussian profile to the resulting image
-- we use the CO(6--5) here since the resolution is slightly higher
than the CO(1--0).  The resulting sizes in the source plane are
r$_{CO}$=110--300\,pc, comparable to the Jeans length computed in \S3.3
for the gaseous ISM in this system, which would be the expected from
the {\it initial} fragmentation scale of a gravitationally-unstable
gaseous disk.  Since the source-plane reconstruction is sensitive to
the lensing model, we also reconstruct and measure the sizes of each of
the star-forming regions from the family of lens models which
adequately reproduce the lensing configuration, which, on average
introduces a $\sim$50\% error on this size measurement.  Since
measuring the sizes for gas emission is difficult at these surface
brightness levels, for all size measurements, we conservatively allow a
factor 2$\times$ error, and report these in Table~2.  Finally, we
estimate the molecular gas mass in the vicinity of the star-forming
regions using the CO(1--0) line luminosity, obtaining
$M=(4$--$15)\times$10$^{8}$\,M$_{\odot}$ for $\alpha$=2.  Together, the
gas in the vicinity of the brightest star-forming regions therefore
comprises approximately 10--20\% of the total gas in the system.

Finally, before investigating the properties of the star-forming
regions in detail, it is useful to compare their characteristics with
those in comparably luminous galaxies where comparable resolution
observations have been made.  In particular, \citet{Downes98}
\citep[see also][]{Sakamoto08} show that the molecular gas within the
central regions of local ULIRGs are characterised by $Q<1$, typically
with two to three compact, dense regions with radii of
$\sim70$--100\,pc and masses $\sim 10^{9}$\,M$_{\odot}$ (and hence
densities, $\sim 2\times 10^4$\,cm$^{-3}$) and typical infrared
luminosities of $(0.3$--$1)\times 10^{12}$\,L$_{\odot}$.  These
properties are similar to those for the clumps in SMM\,J2135.  However,
the key difference is that the bright star-forming regions in local
ULIRGs tend to be located within the central 50--200-pc radius of the
disk, whereas the star-forming regions within SMM\,J2135 are
distributed across $\sim2$\,kpc.  This shows that while the star
formation is occuring in similarly compact clumps, these clumps are
more widely distributed across a much larger gas disk in this
high-redshift galaxy.

\subsection{Cloud scaling relations in a high-pressure environment}
\label{sec:larson}

\begin{figure*}
\centerline{
  \psfig{figure=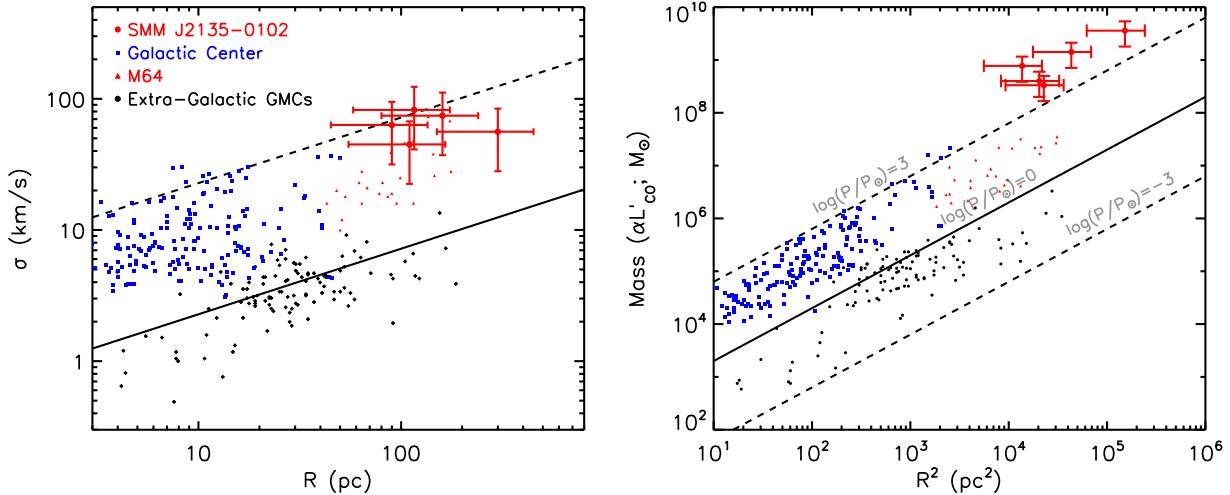,width=6.5in,angle=0}}
\caption{The molecular cloud scaling relations for the star-forming
  regions within SMM\,J2135 compared to those in galaxies in the local
  Universe.  {\it Left:} The velocity dispersion versus cloud
  radius. The extragalactic GMCs in quiescent environments define the
  local line width--size relation (Larson 1981), with the normalisation
  consistent with the average mid-plane pressure in the mid-plane of
  the Milky Way (Elmegreen 1989). Those GMCs in gas-rich,
  high-turbulent-pressure environments (such as the Galactic Center)
  tend to be systematically offset from this relation. The line
  width--size data for star-forming regions in SMM\,J2135 are
  compatible with a line width--size relation with a $\sim 5\times$
  higher normalisation than found in the Milky Way.  {\it Right}: The
  $M({\rm H}_2)-R^2$ relation (equivalent to the second cloud scaling
  relation expressed in Eq.~\ref{eqn:Larson2}). The molecular
  star-forming regions in SMM\,J2135 lie on the line corresponding to
  the very high turbulent pressures estimated for them
  (Equation~\ref{eqn:Ptot}).} 
\label{fig:LRS}
\end{figure*}

Next, we compare the size, luminosity and velocity dispersion
of the star-forming regions to local GMCs and starburst complexes.  For
the line luminosities and velocity dispersions, we use the average from
the  CO(1--0) and CO(6--5) measurements, but ensure that the errors
include the range of solutions from each of the two transitions
separately.  In Fig.~\ref{fig:LRS} we compare the size and velocity
dispersion of the molecular gas associated with the star-forming
regions.  This figure shows that the star-forming regions do not lie on
the local relation for GMCs in quiescent environments
\citep{Bolatto06}, but are instead systematically offset such that at a
fixed size, the velocity dispersion is $\sim 10\times$ higher than
typical GMCs.  Moreover, the offsets of the clumps seen in
Fig.~\ref{fig:LRS} from local scaling relations is comparable to that
found for GMCs in the Galactic Center and other gas-rich environments.  These
offsets have been interpreted as a consequence of the high external ISM
pressure on the cloud surfaces \citep{Keto05, Blitz04, Blitz06}.

\begin{table*}
\begin{center}
{\footnotesize
{\centerline{Table 2: CO properties of the clumps}}
\begin{tabular}{llcccccc}
\hline
\noalign{\smallskip}
                                                                &   Z$_{1}$             & Z$_{2}$            &  Y$_{1}$           &  Y$_{2}$           &   X                & Galaxy Integrated \\
\hline
\\
$S_{\rm 870\mu m}$ (mJy)                                        &  8.2$\,\pm\,$2.1      & 8.2$\,\pm\,$2.1    & 6.4$\,\pm\,$2.1    & 19.1$\,\pm\,$2.1   & 7.5$\,\pm\,$2.1       & 106.0$\,\pm\,$3.5  \\
Amplification ($\mu$)                                           &  63$\,\pm\,$14        & 63$\,\pm\,$14      & 53$\,\pm\,$5       & 40$\,\pm\,$4       & 20$\,\pm\,$2          & 37.5$\,\pm\,$4.5   \\
SFR (M$_{\odot}$\,yr$^{-1}$)                                    &  31$\,\pm\,$10        & 31$\,\pm\,$10      & 30$\,\pm\,$10      & 97$\,\pm\,$14      & 70$\,\pm\,$14         & 400                \\ 
Vel (km\,s$^{-1}$) ($z_{\rm sys}=2.32591$)                      & $-$185$\,\pm\,$19       & $-$63$\,\pm\,$9      & 116$\,\pm\,$12     & 280$\,\pm\,$20     & 380$\pm$20            & ...                \\
$S_{\rm CO(1-0)}$ (Jy\,km\,s$^{-1}$)                            & 0.06$\,\pm\,$0.01     & 0.05$\,\pm\,$0.01  & 0.08$\,\pm\,$0.01  & 0.11$\,\pm\,$0.01  & 0.14$\,\pm\,$0.01     & 2.12$\,\pm\,$0.02  \\
$L'_{\rm CO(1-0)}$ ($\times$10$^8$K\,km\,s$^{-1}$\,pc$^{2}$)    & 2.0$\,\pm\,$0.6       & 1.7$\,\pm\,$0.6    & 3.9$\,\pm\,$0.4    & 7.1$\,\pm\,$0.8    & 1.8$\,\pm\,$0.2       & 173$\,\pm\,$9      \\
$S_{\rm CO(6-5)}$ (Jy\,km\,s$^{-1}$)                            & 0.57$\,\pm\,$0.05     & 1.40$\,\pm\,$0.05  & 1.53$\,\pm\,$0.05  & 1.80$\,\pm\,$0.05  & 1.02$\,\pm\,$0.05      & 21.5$\,\pm\,$1.1   \\
$L'_{\rm CO(6-5)}$ ($\times$10$^8$K\,km\,s$^{-1}$\,pc$^{2}$)    & 0.54$\,\pm\,$0.05     & 1.30$\,\pm\,$0.06  & 1.94$\,\pm\,$0.07  & 3.1$\,\pm\,$0.09   & 3.41$\,\pm\,$0.17     & 48$\,\pm\,$2       \\
$\sigma_{\rm CO(6-5)}$ (km\,s$^{-1}$)                           & 45$\,\pm\,$15         & 82$\,\pm\,$7       & 63$\,\pm\,$12      & 74$\,\pm\,$12      & 56$\pm$15             & 210$\,\pm\,$30     \\
$\sigma_{\rm CO(1-0)}$ (km\,s$^{-1}$)                           & 47$\,\pm\,$20         & 110$\,\pm\,$20     & 57$\,\pm\,$14      & 79$\,\pm\,$15      & 78$\pm$15             & 210$\,\pm\,$30     \\
$r_{\rm CO(6-5)}$ (pc)$^{*}$                                    & 110                   & 116                & 105                & 160                 & 300                    & ...                \\
$r_{\rm 870\mu m}$ (pc)                                         & 90$\,\pm\,$20         & 90$\,\pm\,$20      & 98$\,\pm\,$30      & 180$\,\pm\,$230     & [352,452]$\,\pm\,$50  & ...                \\
M$_{\rm gas CO(1-0)}$ ($\times$10$^8$\,M$_{\odot}$)             & 4.0$\,\pm\,$1.5       & 3.3$\,\pm\,$1.5    & 7.8$\,\pm\,$0.8    & 14.2$\,\pm\,$1.6   & 3.6$\,\pm\,$0.4       &  ...               \\
M$_{\rm gas CO(6-5)}$ ($\times$10$^8$\,M$_{\odot}$)                & 3.8$\,\pm\,$0.4      & 9.2$\,\pm\,$0.8   & 13.8$\,\pm\,$1.0   & 22.8$\,\pm\,$1.0   & 24.3$\,\pm\,$1.6      &  ...               \\
\hline
\end{tabular}
\caption{Notes: the line luminosities, $L'$, are corrected for lensing
  amplification.  The gas mass ($M_{\rm gas}$) is calculated from
  $M_{\rm gas} = \alpha L'_{CO(1-0)}$ with $\alpha=2$ (and
  R$_{61}=$\,0.28\,$\pm$\,0.03 for the CO(6--5);
  \citealt{Danielson11}).  For rows marked with $^{*}$, we estimate
  that the uncertainty on the measurement is a factor $\sim$2.}}
\end{center}
\end{table*}

We can compare the offsets in the size--velocity relation with
theoretical expectations for the line width--size relation for GMCs in
high-pressure ISM environments.  Two important scaling relations
deduced for molecular clouds in the Galaxy are the so-called Larson's
scaling laws \citep{Larson81}, namely: a) the CO velocity
line-width--size relation $\sigma = A R^{a}$, and b) the average gas
density--size relation $\langle n({\rm H}_2)\rangle = B R^{-k}$.  The
first is the most robust of the two, and has been found to be
remarkably constant in the Galaxy, both in terms of its normalisation
and power-law index \citep{Heyer04}.  This reflects the
near-universality of turbulence in molecular clouds, while suggesting
that the energy injection necessary to drive the turbulence occurs
mostly at low spatial frequencies (i.e.\ large scales).  The turbulent
molecular clouds found to obey the Larson relations are almost always
near-virialised \citep{Larson81}; in such a case, the density--size
power law can be readily derived from the line-width--size relation and
the virial theorem.  Indeed following \citet{Elmegreen89} \citep[see
  also][]{Chieze87}, application of the virial theorem on molecular gas
regions of radius, $R$, and boundary pressure, $P_{\rm ext}$ yields

\begin{equation}
  \sigma = \sigma_{\circ} \left(\frac{P_{\rm ext}/k_{\rm B}}{10^4\,{\rm
      K\,cm}^{-3}}\right)^{1/4}\left(\frac{R}{\rm pc}\right)^{1/2}
  \label{eqn:Larson1}
\end{equation}

and

\begin{equation}
  \langle n({\rm H}_2) \rangle = n_0 \left(\frac{P_{\rm ext}/k_{\rm
      B}}{10^4\,{\rm K\,cm}^{-3}}\right)^{1/2} \left(\frac{R}{\rm
    pc}\right)^{-1}
  \label{eqn:Larson2}
\end{equation}

\noindent where $\sigma_{\circ} = 1.2$\,km\,s$^{-1}$ and $n_{\circ}\sim
10^3$\,cm$^{-3}$, as obtained from studies of molecular
clouds in the Galaxy \citep{Larson81, Wolfire03, Heyer04}. 
Theoretical studies predict similar values to within a
factor of two \citep{Chieze87, Elmegreen89}.
Equation~\ref{eqn:Larson2} can be recast as a mass--radius relation
\citep{Elmegreen89} and derived from the velocity--line width relation
and the virial theorem applied for gaseous regions with boundary
pressure.  For the chosen normalisation value of the former,
$n_{\circ}$, this yields:

\begin{equation}
  M({\rm H}_2)=290\left(\frac{P_{\rm ext}/k_{\rm B}}{10^4\,{\rm cm}^{-3}\,{\rm K}}\right)^{1/2}\left(\frac{R}{\rm pc}\right)^2\,{\rm M}_{\odot}.
  \label{eqn:MH2}
\end{equation}

The large mid-plane hydrostatic pressure, $P_{\rm tot}/k_{\rm B}\sim
(1-3)\times 10^7$\,K\,cm$^{-3}$ in SMM\,J2135 will correspond to an
external cloud boundary pressure $P_{\rm ext}\approx P_{\rm
  tot}/(1+\alpha_{\circ}+\beta_{\circ}) = (0.6-1.8)\times
10^7$\,cm$^{-3}$\,K for relative cosmic ray and magnetic pressure
contributions of $\alpha_{\circ}=0.4$ and $\beta_{\circ}=0.25$
\citep[][and references therein]{Elmegreen89}.  From
equations~\ref{eqn:Larson1} \&~\ref{eqn:Larson2} we then expect
$\sigma_{\circ}({\rm SMMJ2135})\sim 7\times \sigma_{\circ}({\rm
  Galactic})$ and $n_{\circ}({\rm SMMJ2135})\sim 50\times
n_{\circ}({\rm Galactic})$.  The former is in very good agreement with
the offset for the star-forming regions (a factor 7--10 offset in
$\sigma$ at a fixed size), whilst the latter implies that the molecular
gas is expected to be $\sim 50\times$ denser {\it on all scales} when
compared to the gaseous ISM in the Milky Way.  Thus for a typical GMC
size of $2 R\sim 20$\,pc in SMM\,J2135, the average densities would be
$\sim 5000$\,cm$^{-3}$ rather than $\sim 100$\,cm$^{-3}$ as in the
Milky Way.

In Fig.~\ref{fig:LRS} we also plot the mass--radius relation for a
range of turbulent ISM pressures and overlay observed values of $M$ and
$R$ from Galactic GMCs, ranging from those in the disk to those in the
high-pressure environment of the Galactic Center.  The star-forming
molecular gas regions of SMM\,J2135 are offset by two orders of
magnitude with respect to quiescent GMCs, as expected from
Equation~\ref{eqn:MH2}, and the turbulent pressures deduced for its
gaseous ISM.  

Whilst our analysis could readily attribute the offsets of the line
width--size relation to the independently-derived large ISM pressures,
we caution that we may still be probing scales at or just beyond where
this relation is valid.  Indeed this scaling law is expected to have a
cut-off at the scales corresponding to the largest possible virialised
molecular cloud structures, or simply to the lowest spatial frequency
of turbulent driving (the disk scale-height being an obvious choice).
The original study by \citet{Larson81} extends up to a cloud size of
100\,pc, while the most recent and systematic study by \citet{Heyer04}
only goes up to 50\,pc (a typical GMC size).  On larger scales,
systematic (streaming) motions between otherwise virialised gas
structures could broaden the apparent ``cloud'' line widths, our
removal of regular disk-like motions notwithstanding.  Such
systematically-broadened line widths could then create apparent offsets
from the local scaling relations, similar to those seen in
Fig.~\ref{fig:LRS}, that are unrelated to increased ISM pressures.
Only ALMA will be capable of verifying the true pressure-induced
offsets in distant star-forming galaxies, by probing scales
$<$\,100\,pc with sufficient sensitivity and spatial resolution to
establish both the slope and the normalisation of the ISM scaling laws
at high redshifts.  This will be possible for strongly-lensed starburst
systems such SMM\,J2135, allowing an unprecedented insight into key
quantities characterising the turbulent molecular gas in star-forming
systems at high redshifts.

\subsection{Star-formation efficiencies at high ISM pressures}

High average volume densities for molecular gas have been revealed by
HCN, CS and HCO$^+$ observations of local ULIRGs and interpreted as
being responsible for the very high star-formation efficiencies
observed for their global molecular gas reservoirs \citep{Greve09}.
  The high luminosities of the
star-forming regions in SMM\,J2135 per gas mass suggest high
efficiencies, with $\Sigma_{\rm FIR}/\Sigma_{\rm gas} =
(100$--$450)$\,L$_{\odot}$/M$_{\odot}$, near the radiation-pressure
limit recently advocated for extreme starbursts (\citealt{Andrews11};
see also \citealt{Scoville04conf} for a first derivation).

However, with estimates for the size, velocity dispersion, mass and
star-formation rates of the clumps in SMM\,J2135, we can estimate the
star-formation efficiencies per dynamical time directly.  Using the
radii of $R\sim 100$--$200$\,pc (as measured from the CO(6--5) line
emission and the rest-frame 260-$\mu$m emission), for velocity
dispersions of 40--60\,km\,s$^{-1}$, the crossing time of the clump
region is $\tau\sim 4$--5\,Myr.  The inferred star-formation rates of
each region is $\sim 30$--90\,M$_{\odot}$\,yr$^{-1}$, thus in one
dynamical time, 1--5$\times$10$^{8}$M$_{\odot}$ of gas will be
converted into stars.  Assuming that all of the molecular gas in a GMC
is converted into stars (i.e.\ no gas loss), it takes 2--10 dynamical
times to convert 10$^{9}$M$_{\odot}$ into stars.  Conversely, the
star-formation efficiency per dynamical time must be $\epsilon_{cl}\sim
10$--50 per cent, i.e.\ approaching the star-formation efficiency of
dense HCN-bright cores in the Milky Way.  This is indeed a natural
outcome of the much higher densities expected at every spatial scale in
the high-pressure ISM within SMM\,J2135 (as discussed in
\S\ref{sec:larson}).  Clearly these calculations are simplistic since
they assume that there is no gas loss and that the star-formation rate
remains constant over the gas depletion lifetime.  Nevertheless, they
demonstrate that the star-formation efficiency in the clumps can be
much higher than those in molecular clouds in quiescent GMCs.  It is
worth noting that the galaxy-averaged star-formation efficiency is much
lower, $\epsilon_{gal}\sim\epsilon_{cl}\times
M_{clumps}/M_{total}\sim2$--10\% (i.e.\ closer to the average value
seen in the Schmidt-Kennicutt relation; \citealt{KS98}).

%
%

\begin{figure}
\centerline{
  \psfig{figure=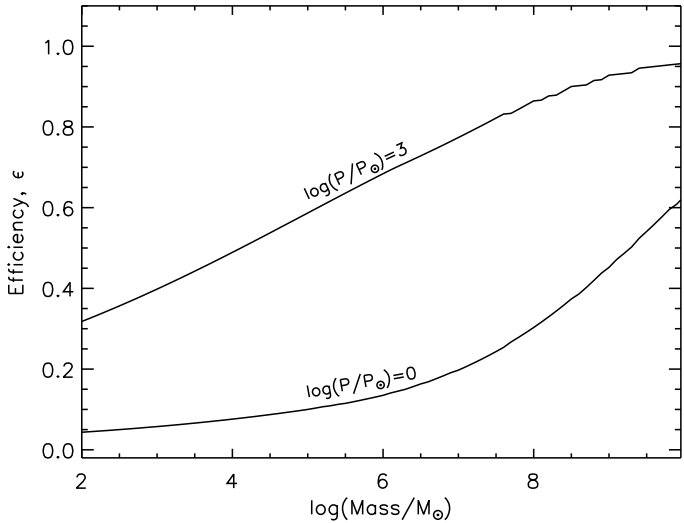,width=3.0in,angle=90}}
\caption{Efficiency of star formation as a function of cloud mass and
  ISM pressure ($P_{\odot}$ denotes the Milky Way value).  The lines
  show the calculation from \citet{Elmegreen97} for the efficiency of
  star formation as a function of the cloud mass in an environment with
  pressure $10^3\times$ that of the Milky Way (the efficiency as a
  function of cloud mass for the Milky Way mid-plane pressure,
  log($P/P_{\odot}=0$) is also shown).  For the ISM pressure measured
  in SMM\,J2135 (log($P/P_{\odot}=3$) and typical GMC masses of $\gsim
  10^5$\,M$_{\odot}$, the star-formation efficiency can be as high as
  $\sim 80$--90 per cent.}
\label{fig:EE97_pressure}
\end{figure}

We can compare these values to a theoretically-derived maximum
efficiency expected for high-pressure molecular gas in star-forming
regions.  \citet{Elmegreen97} suggest that if both the gas density and
the rate of star formation is high enough, most of the gas in a cloud
may be consumed before the molecular cloud is dispersed.
\citet{Elmegreen97} also show that high-pressure environments result in
a lower dispersal rate, which improves the efficiency (resulting from
the higher physical density and gravitational binding energy).  In this
model, the star-formation efficiency depends on the relative rate of
star formation (which is proportional to some power of the gas density,
$n\sim 1.5$; \citealt{KS98}) and the rate of dispersal (which is
proportional to the stellar luminosity and inversely proportional to
the gravitational binding energy).  By scaling the relative rates of
star formation and efficiency from those in the Milky Way,
\citet{Elmegreen97} estimate the star-formation efficiency as a
function of pressure and cloud mass, and in
Fig.~\ref{fig:EE97_pressure} we show how the star-formation efficiency
scales with ISM pressure and mass.  For typical GMC mass scales of
$\gsim 10^5$\,M$_{\odot}$ and external ISM pressures of $10^3\times$
that of the Milky Way, the star-formation efficiency can be as high as
$\sim 80$ per cent.

\section{Implications for star-formation in highly turbulent gas disks}

The large line width observed for the high-pressure molecular gas of
extreme starbursts like SMM\,J2135 will alter the so-called sonic or
turbulent pressure length $\lambda_{\rm s}$.  This marks the scale at
which supersonic turbulence becomes subsonic and near-thermal, and
below which micro-turbulent velocity fields dominate the gas motion
\citep{Wolfire03, Krumholz05}.  In other words $\lambda_{\rm s}$ marks
the smallest scale where the line width--size relation can hold in
molecular clouds, below which there are near-constant velocity
dispersions that no longer correlate with sizes \citep{Plume97}.  Using
a line width--size relation re-cast in terms of the isothermal sound
speed $c_{\rm s}$, namely $\sigma (R)=c_{\rm s}\left(R/\lambda_{\rm
  s}\right)^{1/2}$, it is clear that a $\sim 10\times$ higher
normalisation factor implies a difference in $\lambda_{\rm s}$ between
SMM\,J2135 and the Milky Way of $\lambda_{\rm s}^{\rm SMM}/\lambda_{\rm
  s}^{\rm MW}\sim 0.01$ (assuming the same sound speed).  This means
that in systems such as SMM\,J2135, much smaller gaseous structures can
remain supported against gravitationally-induced fragmentation/collapse
by strong turbulence is dominant on $\sim 100\times$ smaller scales
than in the kinematically quiescent ISM of the Galaxy.

In the context of turbulence-regulated star formation, only regions
with overdensities $x=n/\langle n \rangle$ can ``escape'' the grip of
supersonic turbulence and irreversibly proceed towards gravitational
collapse \citep{Krumholz05}.  For the Milky Way, the critical value is
$x_{crit}=1.07$M$^2$ (where $M$ is the one-dimensional Mach number).
For SMM\,J2135, our results suggest,
$x_{crit}$(SMM\,J2135)$\sim$100$x_{crit}$(Galaxy).  In absolute
densities, for $\langle n \rangle $(SMM\,J2135)$\sim
$10$^{4.1}$\,cm$^{-3}$ \citep{Danielson11}, the critical density will
be $\rm n_{crit,gr}$$\sim $$1.4\times 10^8$\,cm$^{-3}$.  This is an
exceptionaly high density, and for perspective the corresponding
critical density in the Milky Way is $\sim $10$^4$--10$^5$\,cm$^{-3}$.
As such, the normalized star-formation-rate (SFR) per free fall time
will be increased substantially \citep{Padoan11}.  Far from being the
exception, star formation in such highly turbulent ISM will not be
atypical in the distant Universe where starbursts, such as SMM\,J2135,
represent an important star-forming population.  Indeed, half of the
stars seen today may have formed in such environments \citep{Blain99b}.

\subsection{The thermal state for dense highly turbulent gas}

The higher normalization of the linewidth-size relation in the ISM of
SMM\,J2135 translates to a much smaller sonic length controling the
onset of near-thermal motions and eventual gravitational collapse.
However this assumes isothermal gas, and thus an invariant sound speed
$\rm c_s$.  While this has been widely used in numerical simulations
\citep[e.g.][]{Padoan02}, and analytical models of turbulence-regulated
star-formation in individual molecular clouds and gaseous galactic
disks \citep[e.g.][]{Krumholz05}, it does not represent a realistic
ISM.  Indeed, non-isothermal gas may play a crucial role in defining
the thermodynamic state of the dense gas and hence on the stellar IMF
and its mass scale (i.e. the IMF ``knee'';
\citealt{Li03,Jappsen05,Bonnell06})\footnotemark

\footnotetext{The polytropic Effective Equation of State $\rm
  P_{thermal}=K\rho ^{\gamma} $ is a simple way to incorporate the
  complexities of strongly evolving gas heating and cooling processes
  in such numerical simulations and the values of its polytropic index
  $\gamma$ determine the degree of gas fragmentation and the resulting
  dense gas core mass spectrum \citep[e.g.][]{Li03,Jappsen05}.  The
  dependence of $\gamma $ on the properties of the ambient ISM (e.g.
  metallicity, velocity field, background radiation) have been explored
  in some detail \citep{Spaans00}.}

The very strong turbulence present in the ISM of galaxies such as
SMM\,J2135 (as well as local ULIRGs) will influence the thermodynamic
state of their UV-shielded dense gas phase (where the initial
conditions for star formation are set) in two important, yet opposing
ways.  First, it will volumetrically heat this gas phase (i.e.\ unlike
the surface-like heating of far-UV, optical or infrared photons), and
second it will enhance molecular line cooling in much deeper and denser
regions of the ISM.  The latter occurs simply because {\it highly
  supersonic line widths can now be maintained in much smaller, and
  thus denser gas regions} (cf.\ Eq.~\ref{eqn:Larson2}).  This will
reduce the radiative trapping of cooling molecular line photons from
such regions deep inside molecular clouds, as molecular line optical
depths, $\tau\propto (dV/dR)^{-1}$, can now remain low at much higher
densities because of the much larger velocity gradients, $dV/dR$.
These two effects can significantly alter the initial conditions for
star formation in the highly turbulent ISM by changing the dense core
mass spectrum, and/or the density, $n_c$ setting the equation of state
inflection point (the latter is the density where the polytropic index
flips from $<$1 (for $n\leq n_c$) to $>1$ (for $n> n_c$), its value may
determine the IMF mass scale \citep{Larson05}).

%
%

The effects of large velocity gradients on the degree of gas
fragmentation and the resulting mass function of dense cores have been
explored, but only up to the rather low value of $dV/dR\sim
3$\,km\,s$^{-1}$\,pc$^{-1}$ \citep{Spaans00}.  In the case of
SMM\,J2135, where strong turbulence can be maintained at $n({\rm
  H}_2)\ga 10^5$\,cm$^{-3}$, the average velocity gradient expected
from self-gravity-induced gas motions alone will be:

\begin{eqnarray}
  \left(\frac{dV}{dr}\right)_{\rm vir}&\approx& 0.65\sqrt{\alpha}
  \left(\frac{n({\rm H}_2)}{\rm
    10^3\,cm^{-3}}\right)^{1/2}\,\approx6.5\sqrt{\alpha}\,\rm
  km\,s^{-1}\,pc^{-1}\,\,\,\,\,\,\,\,
  \label{eqn:dVdR}
\end{eqnarray}

\noindent
and may approach $dV/dR\sim10$\,km\,s$^{-1}$\,pc$^{-1}$ (see also
\citealt{Greve09}).  To date, such strongly turbulent media have not
been simulated, even for a purportedly starburst-like ISM
\citep[e.g.][]{Klessen07}.  The discovery of the extreme kinematic
properties of the ISM in SMM\,J2315 makes such simulations necessary,
with preliminary results indicating significant implications for the
resulting dense gas fragmentation and its star-formation efficiencies
(Hocuk \& Spaan, in prep).

\section{Conclusions}

We have mapped the distribution of molecular gas in the lensed,
star-forming $z=2.3$ galaxy, SMM\,J2135, through the CO(6--5) and
CO(1--0) emission lines, using the PdBI and EVLA respectively.  After
correcting for lensing, our high-resolution data provide a physical
resolution of $\sim 100$--200\,pc.  The dynamics of both the CO(1--0)
and CO(6--5) line-emitting material convincingly demonstrate for the
first time in a galaxy at $z\gg 1$, that the molecular gas is located
in a large rotating disk, which suggests such structures may be
ubiquitous in high-redshift galaxies.  From the kinematics we derive an
inclination-corrected rotation speed for the disk of $v_{\rm rot} =
320\pm 25$\,km\,s$^{-1}$, $v/\sigma = 3.5\pm 0.2$ and a dynamical mass
of $(6.0\pm 0.5)\times 10^{10}$\,M$_{\odot}$ within a radius of
2.5\,kpc (the spatial extent of the detected emission in CO(1--0)).
The disk is massively unstable, with a toomre parameter, $Q = 0.50\pm
0.15$, which is lower than the compact, gas-rich disks residing in
local ULIRGs where $Q$ is closer to unity \citep{Downes98}.  We also
find the CO(1--0) emission is slightly more spatially extended than
than the higher-$J$ CO lines, as has been seen for other high-redshift
SMGs and ULIRGs \citep{Ivison11EVLA,Riechers11a,Riechers11b}.
Combining with measurements of the stellar mass within this system, we
find that the dynamics of the disk are baryon dominated,
(M$_{gas}$+M$_*$)/M$_{dyn}$=1.1$\pm$0.2, with a molecular gas fraction
of $M_{\rm gas}/(M_{\rm gas}+M_{\star}) = 0.6\pm0.1$, which is
comparable to that found in other starburst galaxies at these epochs
\citep{Tacconi08,Tacconi10,Engel10}.

We use the gas dynamics and surface mass density to explore the ISM
properties, deriving a mid-plane hydrostatic pressure of $P_{\rm
  tot}\sim (2\pm 1)\times 10^{7}$\,K\,cm$^{-3}$, which corresponds to
an external boundary pressure on the GMCs of $P_{\rm ext} =
(0.6$--$1.8)\times 10^7$\,cm$^{-3}$\,K, a factor $\sim $1000 times
higher than the Milky Way.  Using established expressions of the
(linewidth)-scale and (average density)-scale scaling relations known
in the local ISM we deduce that at these high pressures the expected
velocity dispersions and densities of molecular gas regions should be
$7\times$ and $\sim 50\times$ the Galactic values respectively {\it at
  all scales} where supersonic turbulence remains important.  We use
our high-resolution CO cubes to demonstrate that this is indeed the
case for the star-forming regions in SMM\,J2135, which are similar to
the high-pressure ISM environments found in the Galactic Center.

The systematically higher gas densities expected from the higher
normalization of the (average density)-(scale) relation (or its
equivalent (cloud mass)-(radius) relation) are expected to dramatically
elevate the star-formation efficiencies in this system, providing a
natural explanation for the high star-formation-rate densities within
the compact regions seen in the high resolution rest-frame 260$\mu$m
continuum imaging of this galaxy.  We isolate these star-forming
regions using their rest-frame 260-$\mu$m continuum emission and
extract the gas velocity dispersions and luminosities in the
corresponding regions of the disk.  The molecular mass in the vicinity
of each of the star-forming regions is $M_{\rm cl} = (5$--$15)\times
10^8$\,M$_{\odot}$.  Thus, the brightest star-forming regions in the
galaxy make up approximately $\sim 10$ per cent of the total baryonic
mass in the disk (see also \citealt{Genzel10,Elmegreen09}).  In all four regions,
the velocity dispersions of the molecular gas are 40--85\,km\,s$^{-1}$.

We then compare the measured star-formation efficiencies of the clumps
with those from theoretical expectations of high pressure ISM.  Given
their sizes and velocity dispersions, the crossing time of each of the
star-forming regions is $\tau\sim 4$--5\,Myr.  For star-formation rates
of $\sim 30$--90\,M$_{\odot}$\,yr$^{-1}$, each star-forming region will
convert the $\sim 10^{9}$\,M$_{\odot}$ of molecular gas in their
vicinity into stars in just 2--10 dynamical times.  Conversely the
star-formation efficiency per dynamical time must be $\sim 10$--50 per
cent, a factor $\gsim 10\times$ higher than in the GMCs in the Milky
Way.  Nevertheless, this is compatible with the theoretical expectation
for the star-formation efficiency at high-pressure and cloud mass
\citep{Elmegreen97}.

Finally, recasting the linewidth--scale relation in terms of the sound
speed in an isothermal medium, as well as in the context of
turbulence-regulated star formation, we show that for the highly
turbulent ISM of SMM\,J2135 the turbulence will remain dominant,
preventing gravitational collapse for structures that are up to 100
times smaller (and denser) than the corresponding structures in the
Milky Way, resulting in a high critical density of $\sim
$10$^{8}$\,cm$^{-3}$ for the onset of star formation.  
In the context of a more realistic non-isothermal ISM, we argue that
the highly supersonic gas motions maintained at much higher gas
densities in the turbulent ISM may alter the thermal state and the
effective equation of state of the gas phase from which stars form.
Detailed numerical simulations of such highly turbulent ISM are thus
urgently needed to explore its mass fragmentation spectrum, the
star-formation efficiencies, and the possible impact on the emergent
stellar IMF.

Overall, our observations have resolved the molecular gas emission in a
high-redshift star-forming galaxy at $z=2.3$ on 100-200\,pc scales.
The gas is located in a rotating disk which is highly unstable
(Q=0.50$\pm$0.15), and the gas density suggests the ISM should be under
a mid-plane hydrostatic $\sim$10$^3\times$ that of the Milky Way.  We
show that the gas in the vicinity of the star-forming regions is
$\sim$10$\times$ denser than that of typical GMCs in the Milky Way, and
causes them to be significantly offset from scaling relations which
govern the structure of local GMCs.  These offsets imply that
supersonic turbulence will remain dominant on scales $\sim $100$\times$
smaller than in the kinematically quiescent ISM of the Milky Way, while
the molecular gas will be $\sim $50$\times$ denser {\it on all scales.}
Thus, the high star-formation densities seen in SMM\,J2135 are a
consequence of the high gas surface densities and pressures, which
result in a low dispersal rate for the clouds and star-formation
efficiencies $\sim10\times$ greater than those in quiescent
environments.  In the era of ALMA under full science operations, such
observations should become common place, verifying the effects of
pressure-induced offsets from local GMC scaling relations in distant
star-forming galaxies.

\section*{acknowledgments}

We would like to thank the referee for a constructive report which
significantly improved the content and clarity of this work.  AMS, IRS,
RJI and APT gratefully acknowledges support from STFC.  We thank
Richard Bower, Bruce Elmegreen, Natascha Forster-Schreiber, Reinhard
Genzel, Phil Hopkins, Mark Krunholz, Norman Murray, and Linda Tacconi
for advice and conversations, and Jim Geach for volume rendering the
cube and making the PdBI movie.  The observations in this paper were
carried out with the IRAM Plateau de Bure Interferometer and the
Expanded Very Large Array.  PdBI is supported by INSU/CNRS (France),
MPG (Germany) and IGN (Spain).  The EVLA is a facility of NRAO,
operated by AUI, under a cooperative agreement with the NSF.

\bibliographystyle{apj} 
\bibliography{/Users/ams/Projects/ref}

\end{document}